\documentclass{aa}  
\usepackage{graphicx}
\usepackage{times}
\usepackage{amssymb}
\usepackage{amsmath}				%
\usepackage{multirow}
\usepackage{booktabs}
\usepackage{txfonts}
\usepackage{xcolor}
\bibpunct{(}{)}{;}{a}{}{,} 
\usepackage{hyperref}
\hypersetup{
    colorlinks,
    linkcolor={red!100!black},
    citecolor={blue!100!black},
    urlcolor={blue!100!black}
}
\makeatletter
\renewcommand*\aa@pageof{, page \thepage{} of \pageref*{LastPage}}
\makeatother

\begin{document} 
\title{Massive runaway star HD~254577: the pre-supernova binary companion to the progenitor of the supernova remnant IC~443%
}

\author{%
    B. Din\c{c}el       \inst{1}\and
    G. Payl{\i}         \inst{1}\and
    S.~K. Yerli         \inst{2}\and
    A.\ Ankay           \inst{3}\and
    R.\ Neuh\"{a}user   \inst{1}\and
    M.\ Mugrauer        \inst{1}\and
    S.\ Sheth           \inst{1,4}\and
    S. Buder            \inst{5}\and
    S.\ Hüttel          \inst{1}\and\\
    F.\ Edelmann        \inst{1}\and
    K-U. Michel         \inst{1}\and
    J. B\"{a}tz         \inst{1} 
}

\institute{%
$^{1}$ Astrophysikalisches Institut und Universit\"{a}ts-Sternwarte Jena, 07745 Jena, Germany\\
$^{2}$ Orta Do\u{g}u Teknik \"Universitesi, Department of Physics, 06800 Ankara, Turkey\\
$^{3}$ Bo\u{g}azi\c{c}i University, Department of Physics, 34342 \.{I}stanbul, Turkey\\
$^{4}$ Leibniz-Institut für Astrophysik Potsdam (AIP), An der Sternwarte 16, 14482 Potsdam, Germany\\
$^{5}$ Research School of Astronomy \& Astrophysics, Australian National University, Canberra ACT 2611, Australia\\
\email{baha.dincel@uni-jena.de}
}


\abstract
   {}
   {The secondaries of the massive binary systems can be found as runaway stars after being ejected due to the supernova (SN) of the more massive component. 
   We search for such stars inside the supernova remnants (SNRs), where a recent SN is guaranteed to have happened, and the runaway star is expected to be nearby.
   In this paper, we present the massive runaway star HD~254577 as the pre-supernova binary companion to the progenitor of the supernova remnant IC~443 and the neutron star (NS) CXOU~J61705.3$+$222127.} 
   {We performed spectroscopic observations of the runaway star and specified its atmospheric parameters. We also used archival spectroscopic data on neighboring stars. Together with the precise \textit{Gaia} DR3 astrometry and photometry, we identified the possible birth origin of the runaway star, and by isochrone fitting, we determined the progenitor mass. From the \textit{Gaia} DR3 proper motions, we specified the possible explosion sites and calculated the neutron star (NS) velocity.}
   {HD~254577 is a hot and evolved star with an effective temperature of $24000\pm1000$ K (B0.5II) and a surface gravity of $\log({g~\rm[cm/s^2]})=2.75\pm0.25$. 
It is probably a single star with a peculiar 3-D velocity of $31.3^{+1.2}_{-0.9}$ km~s$^{-1}$, lying at a heliocentric distance of $1701^{+55}_{-54}$ pc.
The cometary tail of the NS implies that it is moving away from the same site as the runaway star.
From the flight trajectories, we calculated that the NS has typical pulsar velocities such as $254-539$ km~s$^{-1}$ at a distance of $1.7$ kpc for $10-20$ kyr ages. Together with the blue-only shifted interstellar medium lines on its spectrum, HD~254577 must be the pre-supernova binary companion to the progenitor of IC~443. By identifying the pre-supernova companion and the possible parent cluster, 
we showed that the progenitor zero-age main sequence mass is high: $31-64$ M$_\odot$, favoring the jet scenario previously proposed.
The SNR distance is precisely determined as $1701^{+55}_{-54}$ pc. 
We also discuss the expansion dynamics of the SNR due to the highly off-centered explosion site. 
The pre-SN binary parameters that we calculated might not favor a strongly interacting binary. 
    }
    {}

\titlerunning{Massive runaway star inside IC~443}
\authorrunning{B.Din\c{c}el et. al. }
\maketitle

\section{Introduction}

A significant portion of stars gain high velocities either by
dynamical ejection due to gravitational interactions of massive stars in cluster cores
\citep{1967BOTT....4...86P}
or a binary disruption resulting from a supernova (SN) explosion of the initially more massive component \citep{1961BAN....15..265B}.
A kinematical link of a runaway star to a neutron star (NS) like PSR1706$-$16 - $\zeta$ Oph \citep{2020MNRAS.498..899N}, and PSR~J0826$+$2637 - HIP~13962 \citep{2014MNRAS.438.3587T}
, or to a visible SN remnant (SNR) \citep{2015MNRAS.448.3196D} 
can be evidence for a binary supernova scenario (BSS).

In the cases where the compact object does not receive a significant kick and less than half of the total mass is ejected during the SN, e.g., the mass is stored in the secondary by conservative mass transfer, binary disruption does not occur \citep{1997ApJ...483..399V, 1993SSRv...66..309V}.
The runaway high-mass X-ray binaries 4U1700-37 \citep{2001A&A...370..170A}, Vela X-1 \citep{1997ApJ...475L..37K},
and 4U~2206+54 
\citep{2022MNRAS.511.4123H} 
are good examples of such survived binaries.

The asymmetric SN explosion of a sufficiently massive star ($>10$ M$_\odot$) can eject the NS with velocities of 250--1300 km~s$^{-1}$ \citep{2024arXiv240113817J}, which is remarkably faster than most of the objects in the Galaxy.
The measured 2-D pulsar velocities (typically 300--500 km~s$^{-1}$) confirm these theoretical values 
\citep{%
		1994Natur.369..127L,
		1997ARep...41..257A,
		2005MNRAS.360..974H}.
Therefore, most pulsars are found in isolation, as well as the majority of OB-type runaway stars \citep{1996ApJ...461..357S}. Considering the selection effects and the low rate of X-ray binaries, 
the binary disruption is highly likely \citep{2005Ap.....48..330G}.
Binary disruption kinematics is extensively discussed in \cite{1998A&A...330.1047T}.
According to \cite{2019A&A...624A..66R}, $78^{+9}_{-22}$\% of all massive star binaries do not merge before the SN, and $86^{+10}_{-22}$\% of them become unbound due to the SN.
\begin{figure*}[t]
\centering
\includegraphics[width=\columnwidth]{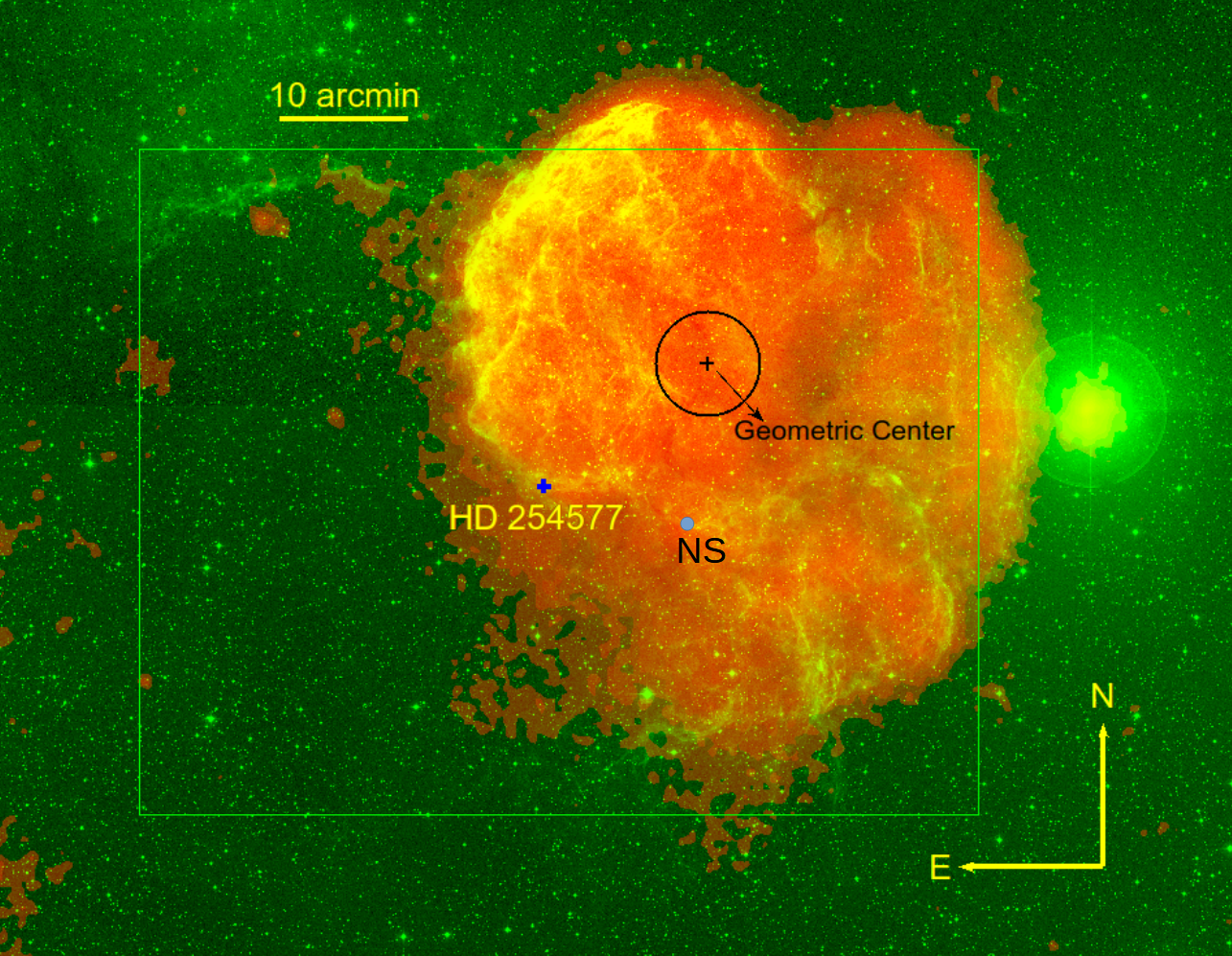}
\includegraphics[width=\columnwidth]{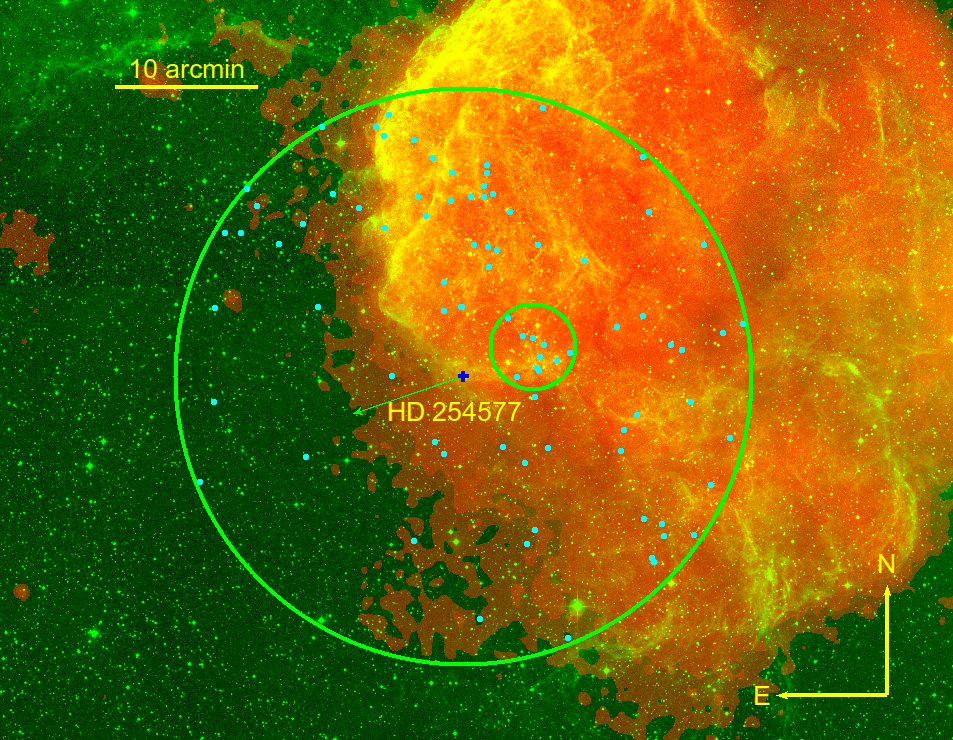}
\caption{\label{f_g189} X-ray (Röntgensatellit (ROSAT) Position Sensitive Proportional Counter (PSPC), 0.1-2.4 keV, in red), and optical (The Space Telescope Science Institute (STScI) 2nd Digitized Sky Survey (DSS2) red image, in green) composite image of IC~443. \textit{Left}: The geometrical center of the SNR is shown with a black cross and the runaway star search region in \cite{2024MNRAS.531.4212D} with a black circle around it. A blue cross indicates the position of HD~254577, while a blue circle indicates that of the NS.
The green box is the border of the image on the right. \textit{Right}: The close-up image of the same. The big green circle represents a region with a radius of 10 pc, hosting a possible stellar group, while the small green circle shows a smaller, more compact group with a 1.5 pc radius. A 100 kyr magnified proper motion vector of HD~254577 is shown with a green arrow. }
\end{figure*}

Finding runaway stars as pre-SN binary companions inside SNRs provides valuable information.
Firstly, the explosion center can be identified more accurately. Thus, the expansion of the SNR and the kick vector of the NS can be studied.
Secondly, the mass of the progenitor star and the pre-SN binary system parameters can be determined, which helps to clarify the origin of certain types of NSs and SNRs. Lastly, SNR distances can be measured precisely using the astrometric or spectroscopic parallax of the runaway star, as its velocity is much slower than the SNR shock wave and stays close to the explosion center in the observational lifetime of the SNR.

Despite the former efforts to find the pre-SN binary companion inside the SNRs \citep{1980JApA....1...67V,
2017A&A...606A..14B,
2019A&A...623A..34K,
2021MNRAS.507.5832K,
2021AN....342..553L,
2024MNRAS.531.4212D}, 
the only highly confident example is the massive runaway star HD~37424 inside SNR S147. It was shown that the runaway star and the pulsar PSR~J0538$+$2817 were once at the same position in space at the same time 30$\pm$4 kyr ago \citep{2015MNRAS.448.3196D}.

We expect that the runaway star in BSS should be young, corresponding to the age of the SN progenitor, exhibit a high peculiar velocity
and either be moving away from the central region of an SNR \citep{2015MNRAS.448.3196D}, share a common origin with an NS \citep{2020MNRAS.498..899N}, or be moving away from a young open cluster where dynamical ejection is no longer expected \citep{2024A&A...691A..63D}. 
The velocity threshold for runaway stars has been discussed in various works. 
$v_{\mathrm{pec}}=28$ and $25$ km\,s$^{-1}$ were suggested as the intersection of the Maxwellian distributions 
of the high and low velocity stellar populations \citep{2011MNRAS.410..190T, Tetzlaff2013}. Significantly higher
values, 40 and 43 km\,s$^{-1}$ were adopted by \cite{2001ApJ...555..364B} and \cite{2024ApJS..272...45G}
respectively, corresponding to the velocity at which the number of stars reaches 1\% of the Maxwellian peak. 
On the other hand, a more conventional value of 30 km\,s$^{-1}$ has often been used as a typical cutoff \citep[e.g.][]{1986ApJS...61..419G,2001A&A...365...49H,2011MNRAS.414.3501E,2019A&A...624A..66R}

A more detailed review of the matter can be found in \cite{2024MNRAS.531.4212D}.
In this paper, we propose that the star HD~254577, inside SNR IC~443, is a 
massive runaway star and a strong candidate for the pre-SN binary companion to the progenitor of SNR IC~443 and the NS CXOU~J61705.3$+$222127. By finding this star, we determined the progenitor mass, the possible pre-SN binary parameters, the possible NS velocities, the age, and the distance of the SNR. 
\section{IC~443 \& CXOU~J61705.3+222127}

IC~443 is a mixed morphology SNR (MMSNR) \citep{1998ApJ...503L.167R} with an angular diameter of 45 arcmin located in the Galactic anticenter direction \citep{2009yCat.7253....0G}. 
The bright shell of the SNR (Fig. \ref{f_g189}) has a similar morphology in radio, X-ray, and optical wavebands
\citep{2004AJ....127.2277L,
1994A&A...284..573A, 
1984ApJ...281..658F}.
The centrally peaked morphology in X-rays \citep{1988ApJ...335..215P} is distinguished from other MMSNRs with peaks at two different positions, 
in the north and the center \citep{2008A&A...485..777T}, explained by cloud evaporation.
The blast wave is in interaction with the surrounding dense atomic and molecular clouds \citep{1997ApJ...489..143C,
2005ApJ...620..758S}.
The SNR has two semispheres (subshells) sharing the same horizontal plane but having different radii and centroids (Fig. \ref{f_g189}) due to expansion in a highly inhomogeneous environment.
From radio observations, \cite{1986A&A...164..193B} concluded that the shock wave is also expanding into a previously generated subshell in the east. Based on X-ray observations, first \cite{1994A&A...284..573A}, with ASKA, and more recently \cite{2023A&A...680A..83C}, with SRG/eROSITA, showed that IC~443 is overlapping with an older SNR G189.6$+$03.3.

The velocities of the shocked optical filaments are in the range of $60-100$ km~s$^{-1}$ \citep{1980ApJ...242.1023F,
1969SvA....13..192L,ba2024},
while the H$_\alpha$ emitting diffuse gas shows velocities up to 350 km~s$^{-1}$.
Considering projection effects, the shock velocity of 400 km~s$^{-1}$ was suggested by \cite{1979A&A....71...29L}.

The SNR is thought to be connected to the Gem OB1 association \citep{1978ApJS...38..309H} 
and Sh~2--249 H$~\textsc{ii}$ region \citep{1968AJ.....73..135G} at $\sim$1.5 kpc; hence, the distance of the SNR has been assumed to be the same.
At this distance, \cite{1988ApJ...335..215P} suggested an age of 3 kyr based on the 
high plasma temperature (10$^7$ K) in the NE rim, while a 30 kyr age was suggested by \cite{1999ApJ...511..798C} from the optical filament expansion and taking the pre-SN bubble into account. 
An age of $4-10$ kyr was derived based on the possible expansion of the ejecta ring and the comparison of the radius of the southern shell \citep{2008A&A...485..777T}.
\cite{2005ApJ...631..935K} proposed a larger age of $10-20$ kyr considering its mixed morphology nature.

Based on NIR Fe$~\textsc{ii}$ line intensities, the extinction toward the SNR largely varies $A_\mathrm{V}=2-6$ mag within the northeastern shell \citep{2013ApJ...768L...8K}. 
A similar result, $A_\mathrm{V}=2.5-3.4$ mag was found 
from H$_\beta/$H$_\alpha$ strength ratios \citep{1984ApJ...281..658F},
which is consistent with the $N_\mathrm{H}\sim~$6.2$\times$10$^{21}$ cm$^{-2}$ 
measured from the X-ray spectral fit \citep{2008A&A...485..777T}.
Due to dense molecular clouds
in which the SNR is expanding, 
the $A_\mathrm{V}$ value increases sharply beyond
$\sim1.6$ kpc in that direction
\citep{2024MNRAS.531.4212D}.

Taking into account the abundances of $\alpha-$process elements detected, 
IC~443 originated from a core-collapse SN \citep{2008A&A...485..777T}.
By studying a jet-like structure in the northwest, 
\cite{2018A&A...615A.157G} proposed a massive progenitor with a mass of $\sim30$ M$_\odot$. 
\begin{table*}[t]
\centering
\caption{\label{clustermembers11}Identified members of the detected potential cluster. \textit{\textit{Gaia}} DR3 names, geometric distances, parallaxes, G-band brightnesses, G$_{BP}-$G$_{RP}$ color indices, and proper motion values are given. \textit{\textit{Gaia}} DR3 3377009162805147136 (the last row) is HD~254577, and \textit{Gaia} DR3 3377012731919116416 (the second row) is HD~254477.
}
\begin{tabular}{@{}l c c c c c c@{}}
\toprule
\textit{Gaia} DR3 Source	    
& r$_{geo}$ (pc)        
& $\varpi$ (mas)  	      
& G (mag) 	     
& G$_{BP}-$G$_{RP}$ (mag) 
&$\mu_\alpha^\star$ (mas~yr$^{-1}$) 	
& $\mu_\delta$ (mas~yr$^{-1}$)  \\
\midrule
3377012938077937280	&$1731^{+115}_{-97}$   &$0.5383\pm0.0342$  &$15.360\pm0.003$    &$1.695\pm0.010$     &$0.557\pm0.038$   &$-1.832\pm0.028$  \\[3pt]
3377012731919116416	&$1579^{+64}_{-54}$    &$0.5998\pm0.0234$  &$10.168\pm0.003$    &$0.908\pm0.028$     &$0.483\pm0.029$   &$-1.813\pm0.021$  \\[3pt]
3377013148534813056	&$1699^{+85}_{-68}$    &$0.5503\pm0.0292$  &$15.161\pm0.003$    &$1.042\pm0.007$     &$0.496\pm0.033$   &$-2.280\pm0.025$  \\[3pt]
3377012697559377664	&$1704^{+42}_{-37}$    &$0.5553\pm0.0139$  &$12.514\pm0.003$    &$0.750\pm0.007$     &$0.376\pm0.016$   &$-2.221\pm0.012$  \\[3pt]
3377011911584237824	&$1716^{+48}_{-43}$    &$0.5403\pm0.0152$  &$13.240\pm0.003$    &$0.820\pm0.007$     &$0.436\pm0.017$   &$-1.564\pm0.014$  \\[3pt]
3377012701858219392	&$1637^{+92}_{-70}$    &$0.5713\pm0.0265$  &$14.910\pm0.003$    &$0.952\pm0.007$     &$0.416\pm0.031$   &$-1.759\pm0.023$  \\[3pt]
3377013182894545536	&$1767^{+59}_{-61}$    &$0.5269\pm0.0179$  &$13.892\pm0.003$    &$0.804\pm0.007$     &$0.351\pm0.020$   &$-1.714\pm0.015$  \\[3pt]
3377013178596140800	&$1827^{+160}_{-145}$  &$0.5036\pm0.0443$  &$15.566\pm0.003$    &$1.379\pm0.009$     &$0.361\pm0.042$   &$-1.713\pm0.033$  \\[3pt]
3377011877224494720	&$1730^{+111}_{-87}$   &$0.5326\pm0.0328$  &$15.383\pm0.003$    &$1.355\pm0.008$     &$0.540\pm0.035$   &$-1.534\pm0.027$  \\[3pt]
3377013316036365440	&$1702^{+70}_{-65}$    &$0.5468\pm0.0303$  &$14.634\pm0.003$    &$0.900\pm0.008$     &$0.352\pm0.029$   &$-1.710\pm0.024$  \\[3pt]
3377009162805147136	&$1701^{+55}_{-54}$    &$0.5513\pm0.0175$  &$ 8.761\pm0.003$    &$1.182\pm0.009$     &$4.303\pm0.019$   &$-1.464\pm0.014$  \\
\bottomrule
\end{tabular}
\end{table*}

The compact object of the SNR is a radio-quiet X-ray NS CXOU~J61705.3$+$222127.
Its pulsar wind nebula (PWN) and blackbody radiation are observable in X-rays 
\citep{2001ApJ...554L.205O, 2001A&A...376..248B}.
So far, no pulse has been detected, probably because the polar beams are not sweeping our line of sight.
The hydrogen atmosphere model yields an effective temperature of $T_\mathrm{eff}\approx6.8\times10^{5}$ K, while the blackbody model results in a higher temperature, 
$T_\mathrm{eff}\approx1.6\times10^{6}$ K, which is close to that of the Vela pulsar
\citep{2015ApJ...808...84S}, a $10-30$ kyr old NS.
The cooling age is also consistent with the SNR age of 30 kyr \citep{2006ARA&A..44...17G}.
The pulsar period and the magnetic field strength are predicted to be $P=250$ ms
and $B_\mathrm{dip}=2.0\times10^{13}$ G, respectively, on the basis of the X-ray bolometric luminosity of the PWN, which is $L_\mathrm{PWN}\approx1.4\times10^{33}$ erg~s$^{-1}$.
CXOU~J61705.3$+$222127 is an energetic pulsar with a possible spin-down luminosity of $\dot E=(1-30)\times10^{36}$ erg~s$^{-1}$.
The proper motion of the NS was measured with a high uncertainty;
$\mu^*_\alpha=-22.3\pm33.9$, $\mu_\delta=-0.1\pm33.9$ mas~yr$^{-1}$
\citep{2015ApJ...808...84S}, implying motion mainly towards the West.
From the cometary tail of the PWN in X-rays, the velocity of the NS is estimated to be 230 km~s$^{-1}$ \citep{2006ApJ...648.1037G}.

The position of the NS is $\sim$12 arcminutes south of the geometric center.
However, the bow-shock PWN, which is shaped by the supersonic motion of the NS,
does not point back to the geometric center as its origin \citep{2001ApJ...554L.205O}.
The parallactic angle of the cometary tail is measured as 71$\pm$5$^\circ$, suggesting a east-to-west motion, rather than north-to-south.
According to \cite{2004AJ....127.2277L}, the NS might originate from another SNR, possibly G189.6$+$03.3. 
However, the very soft X-ray spectrum of this remnant, with a mean temperature of $kT\sim0.14$ keV \citep{2004AJ....127.2277L}, implies an age of $\sim100$ kyr \citep{ba2024}, significantly older than both IC~443 and the NS age, inferred from cooling curves. We therefore assume that the NS is the compact remnant of IC~443.

Previously, \cite{2024MNRAS.531.4212D} searched for OB-type runaway stars within the central region of the SNR (Fig. \ref{f_g189}). However, no OB-type runaway star candidate based on the proper motion was found in this central region. On the other hand, when we performed a search for OB-type stars with spectroscopic confirmation from \cite{2013yCat....1.2023S} in the entire SNR, we found that HD~254577 shows a significantly distinct proper motion value relative to the other seven OB-type stars, and might have the exact origin as the NS.
\section{HD~254577 and Neighboring Stars}
HD~254577 is a B0.5II-IIIk type star \citep{1955ApJ...121...24C} located on the western edge of IC~443 (Fig. \ref{f_g189}). It is a pulsating star with a period of $3.991\pm0.002$ days and an amplitude of $6.16\pm0.06$ mmag \citep{2023ApJS..265...33S}. 
The temperature and luminosity measured in \cite{2023ApJS..265...33S} 
are 24332 K and 4.18 L$_\odot$ (no error was mentioned).
The \textit{Gaia} DR3 \citep{2016A&A...595A...1G, 2023A&A...674A...1G} 
parallax of the star is $\varpi=0.5513\pm0.0175$, and the geometric distance is $\mathrm{r_{geo}}=1701^{+55}_{-54}$ pc \citep{2021AJ....161..147B}.
The proper motion of HD~254577 in R.A. and Dec. is $\mu_{\alpha}^*=4.303\pm0.019$ mas~yr$^{-1}$, and $\mu_{\delta}=-1.464\pm0.014$ mas~yr$^{-1}$, respectively.

To determine its peculiar velocity and stellar parameters, 
as well as the progenitor mass, we tried to find a possible parent cluster or stars
that are genetically connected to HD~254577. 
Using the \textit{Gaia} DR3 catalog, we selected stars brighter than $G<17$ mag,
having parallax values between 0.5 and 0.6 mas, detected
on a significance larger than 10, within 20 arcminutes of the runaway star. 
This radius corresponds to $\sim10$ pc at 1.7 kpc, 
a typical tidal radius of an open cluster. 
We found 85 stars (Table \ref{clustermembers}) 
with a very narrow proper motion dispersion, 0.155 and 0.361 mas~yr$^{-1}$ in R.A and Dec., translating to 1.25 and 2.91 km~s$^{-1}$ at 1.7 kpc, respectively.
The average proper motion of this group is then $\mu_{\alpha}^*=0.401\pm0.155$ mas~yr$^{-1}$, and $\mu_{\delta}=-1.775\pm0.361$ mas~yr$^{-1}$ where the errors are the standard deviations.
The distance of this group is $1731\pm68$ pc.
The proper motion of HD~254577 is highly different from that of the group and points out of a more compact group of 10 stars (Fig. \ref{f_g189}), 
a potential young open cluster with a radius of 1.5 pc.
The brightest star in this group is HD~254477, known to be a B8-type star \citep{2002A&A...386..709F}. 
The proper motion of this group is $\mu_{\alpha}^*=0.437\pm0.078$ mas~yr$^{-1}$, 
and $\mu_{\delta}=-1.814\pm0.0.249$ mas~yr$^{-1}$, 
and the distance is $1710 \pm 71$ pc,
well consistent with the distance of HD~254577, $1701^{+55}_{-54}$ pc.
The \textit{Gaia} DR3 names, geometrical distances, parallaxes, brightness in the G-band, color indices, and proper motions are given in Table \ref{clustermembers11}.
The proper motion of the star with respect to the cluster is 
$\mu_{\alpha}^*=3.866\pm0.097$ mas~yr$^{-1}$, and $\mu_{\delta}=0.350\pm0.263$ mas~yr$^{-1}$.

An independent way of calculating the peculiar velocity of the star is to correct its proper motion for Galactic rotation and solar motion.
The Galactocentric distance to the Sun is taken as 8.5 kpc and the solar rotational velocity as 220 km~s$^{-1}$.
We take the local standard of rest from \cite{2011MNRAS.410..190T} as
(U$_{\odot}$,V$_{\odot}$,W$_{\odot}$) = (10.4$\pm$0.4, 11.6$\pm$0.2, 6.1$\pm$0.2) km~s$^{-1}$.
The resultant values for HD~254577 are
$\mu_{\alpha}^*=4.303\pm0.019$ mas~yr$^{-1}$, and
$\mu_{\delta}=-1.464\pm0.014$ mas~yr$^{-1}$

The transverse peculiar velocities are then $31.3^{+1.0}_{-0.9}$ km~s$^{-1}$ w.r.t
the cluster, and $36.6\pm0.2$ km~s$^{-1}$ 
according to the galactic flat rotation model.
In either case, HD~254577 is a runaway star.
\section{Observations}
We performed observations for the runaway star HD~254577 and HD~254477 
to measure the radial velocity (RV) and the atmospheric parameters.
We also used archival spectroscopic data of neighboring stars to determine the temperature and surface gravity, 
and of the SNR filaments to find the real extension of the SNR.
\begin{table}
\centering
\caption{\label{obss}Log of observations for HD~254577 and HD~254477. The observation dates are given together with the spectrograph used, the preferred binning, the detector integration time (DIT) in seconds, and the signal-to-noise ratio obtained around the H$\beta$ $\lambda 4860$ line. The number of DITs is three (NDIT=3) for each observation.}
\begin{tabular}{c c c c c}
\toprule
Date 
& Instrument  
& Binning  
& DIT
& S/N   \\
\midrule
HD~254577  & & & &  \\
\midrule
26 Feb 2015 &FLECHAS &1x1 &1200 &25   \\
16 Mar 2015 &FLECHAS &1x1 &1200 &30   \\
18 Mar 2015 &FLECHAS &1x4 &1800 &65   \\
27 Oct 2015 &HDS     &1x1 &600  &65   \\
29 Jan 2024 &FLECHAS &2x2 &2400 &35   \\
28 Feb 2024 &FLECHAS &2x2 &2400 &40   \\
09 Mar 2024 &FLECHAS &1x1 &2400 &10   \\
25 Mar 2024 &FLECHAS &1x1 &2400 &25   \\
\midrule
HD~254477  & & & & \\
\midrule
22 Feb 2015 &FLECHAS &1x1 &1200 &25    \\
09 Apr 2015 &FLECHAS &1x1 &1200 &20    \\
\bottomrule
\end{tabular}
\end{table}

\subsection{Stellar Spectra}
The spectra of HD~254577 and HD~254477 were taken in several observing epochs with 
the Fibre Linked \' Echelle Astronomical Spectrograph \citep[FLECHAS,][]{2014AN....335..417M}
on the 90 cm Schmidt telescope at the University Observatory Jena (Table \ref{obss}). 
FLECHAS provides a resolving power of R$\sim$9200 with 1$\times$1 pixel binning 
and a wavelength range of 3900-8100 \AA.
The spectra of HD254577 were also taken with 2$\times$2 pixel binning 
which reduces the resolving power to R$\sim$7000. 
For optimal calibration, three arc- (Th-Ar lamp) and flatfield-images
(Tungsten lamp) were taken immediately before the science exposures.
The detector integration time (DIT) of lamp spectra is 5s for 1$\times$1 binning
and 1.5s for 2$\times$2 binning.
The CCD of FLECHAS was operated at a cooling temperature between $-30^\circ$C and $-40^\circ$C.
A set of three dark frames was taken for each DIT used for the arc-,
flatfield-, and science exposures and were median-combined for efficient cosmics removal. 

The high-resolution and high signal-to-noise spectrum of HD~254577 was taken 
with the High Dispersion Spectrograph (HDS) mounted on the Subaru 8.3-m telescope. The configuration served an \' Echelle spectrum with a wavelength range of 3560-4830 {\AA} and 5010-6240 {\AA}, and a resolving power of R$\sim$90000 for the  1$\times$1 binning. A 600s DIT was given for each frame (Table \ref{obss}). Again, three frames were combined to remove cosmics. Five arc- (Th-Ar lamp) and flatfield-images (Tungsten lamp) were taken together with ten bias frames at the end of the night.

The data obtained were reduced in the {\textsc IRAF} environment \citep{1986SPIE..627..733T}.
Dark frames were combined to master darks of all used integration times
that were then subtracted from the science-, flat- and arc frames.
The three frames of each were combined with the average sigma
clipping rejection.
Afterward, the flats were extracted first, then
the science and arcs were done by tracing each \' echelle apertures of the flats.
After extraction, each science was divided by the corresponding flat.
All science spectra were wavelength-calibrated using
the emission lines of the extracted arc spectrum after they were identified. 
Next, the science spectra were normalized and
finally, spectra in all of the apertures were trimmed and combined to have one single spectrum.

Large sky Area Multi-Object fiber Spectroscopic Telescope (LAMOST) \citep{Cui2012}, 
also known as the Guo Shoujing Telescope, located at the Xinglong Station of the National Astronomical Observatories of China (NAOC),
is an astronomical instrument designed for wide-field (5 deg) 
multi-object spectroscopic surveys. 
It covers a wavelength range of 3700-9000 {\AA} in the optical spectrum with a spectral resolution of  
$\mathrm{R}\sim1800$ (low resolution) 
and $\mathrm{R}\sim7500$ (medium resolution).
LAMOST low-resolution spectra of 17 stars, including HD~254477, were also used. We retrieved the spectra from the LAMOST website%
\footnote{http://www.lamost.org/dr9/}.
We then normalized the spectra in the IRAF environment.
\begin{table}                                                             
\centering
\caption{\label{rav}Radial velocity measurements of HD~254577.}
\begin{tabular}{c c c c}
\toprule
Date  
& MJD
& Instrument
& RV (km~s$^{-1}$)\\
\midrule
26 Feb 2015 &57079 &FLECHAS &$16.0\pm8.5$   \\
16 Mar 2015 &57097 &FLECHAS &$15.9\pm8.9$  \\
18 Mar 2015 &57099 &FLECHAS &$15.6\pm5.4$  \\
27 Oct 2015 &57322 &HDS     &$15.5\pm8.8$  \\
29 Jan 2024 &60338 &FLECHAS &$20.4\pm5.0$  \\
28 Feb 2024 &60368 &FLECHAS &$15.7\pm8.2$  \\
09 Mar 2024 &60378 &FLECHAS &$25.2\pm8.5$  \\
25 Mar 2024 &60394 &FLECHAS &$12.4\pm6.4$  \\
\bottomrule
\end{tabular}
\end{table}
\begin{table}
\centering
\caption{\label{sp577}Measured parameters of HD~254577. We determined the same spectral type (SpT\_tw) as \citet{1955ApJ...121...24C} (SpT).}
\begin{tabular}{@{}c c c c c@{}}
\toprule
Stars 
& $T_{\rm{eff}}$ (K)  
& $\log(g)$ 
& SpT\_tw 
& SpT \\
\midrule
HD~254577 &24000$\pm1000$      & 2.75$\pm0.25$ &B0.5II &B0.5II \\
\bottomrule
\end{tabular}
\end{table}

\begin{figure*}
\centering
\includegraphics[height=5.3 cm]{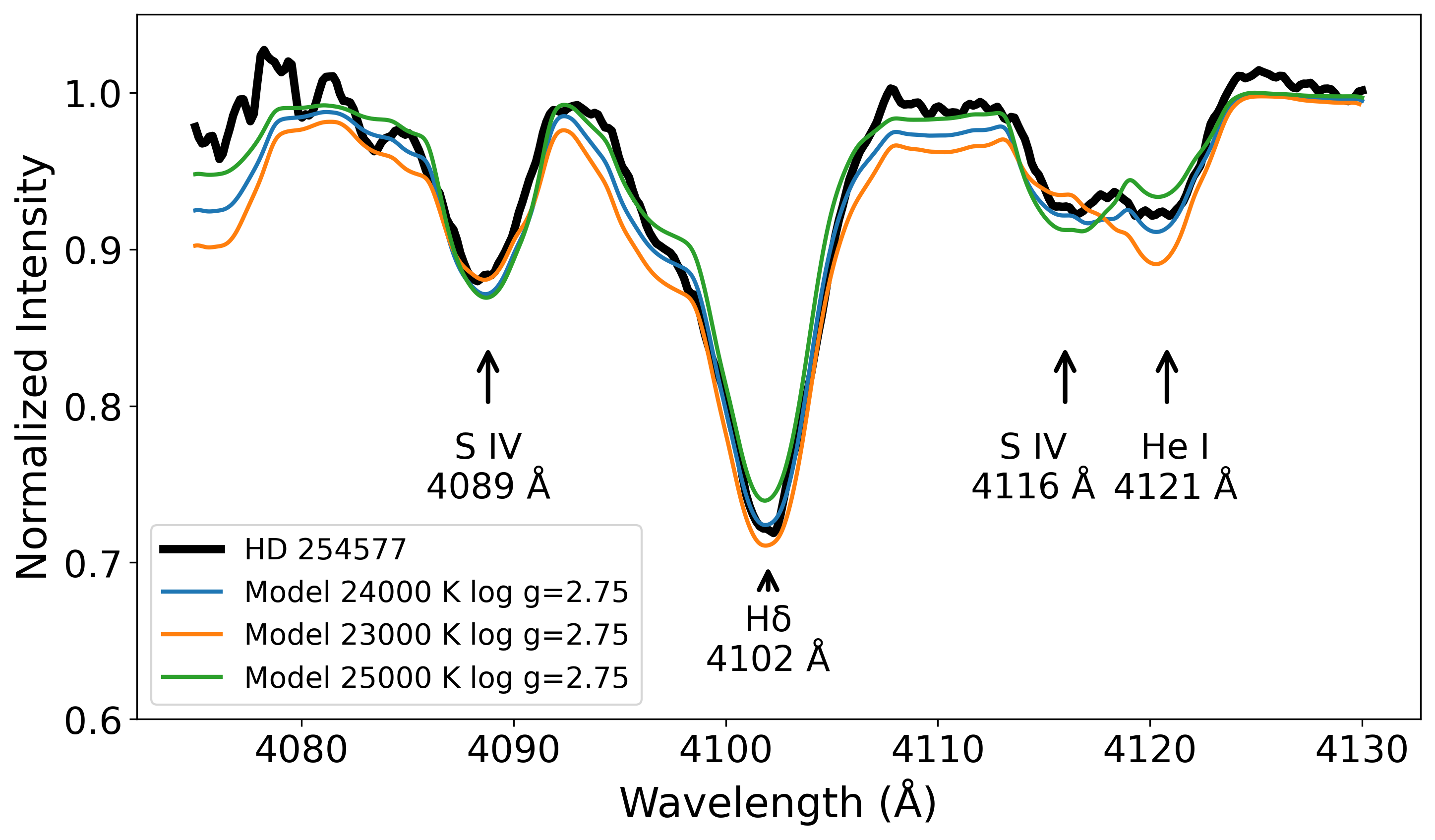}
\includegraphics[height=5.3 cm]{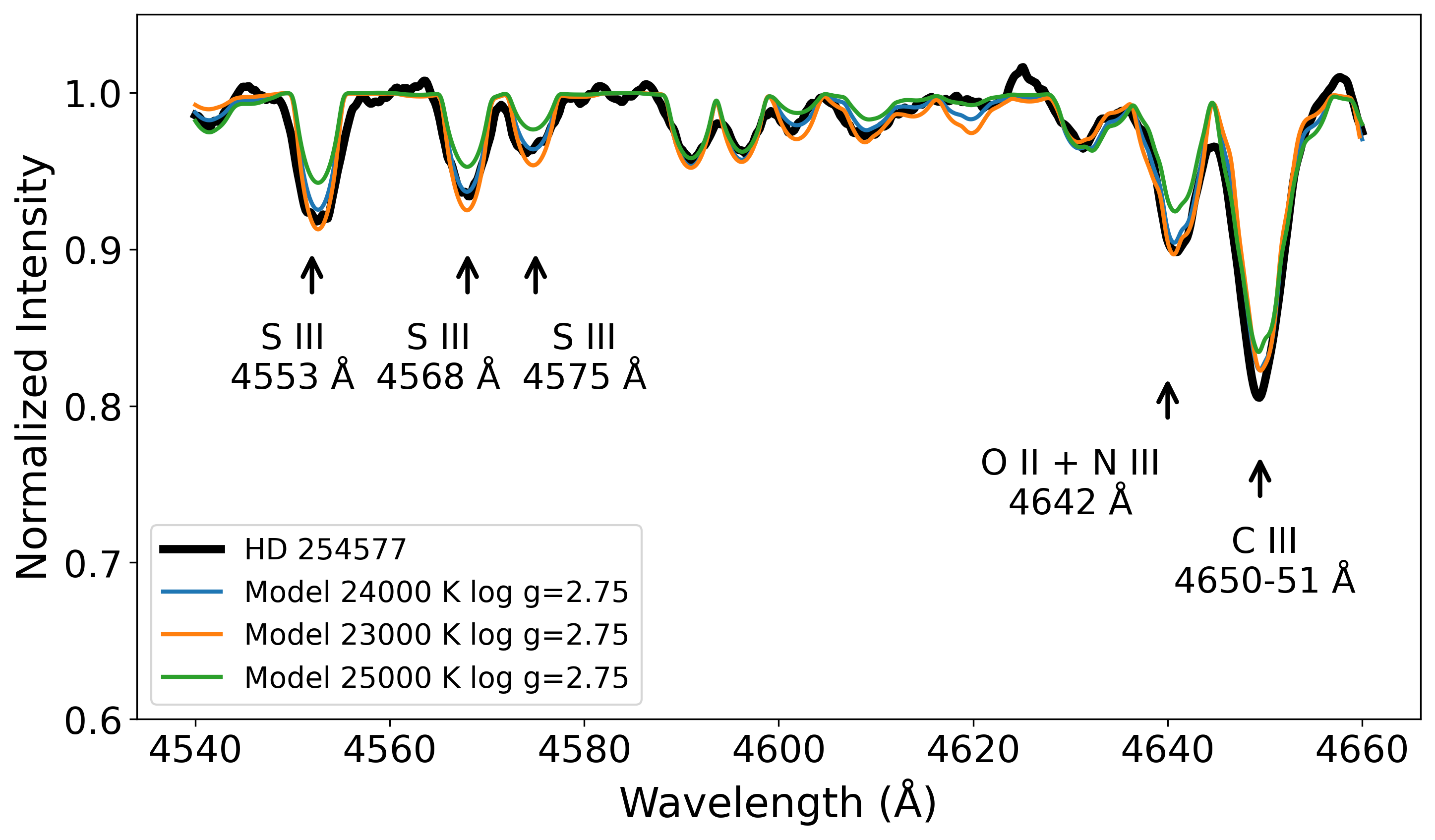}\\
\includegraphics[height=5.3 cm]{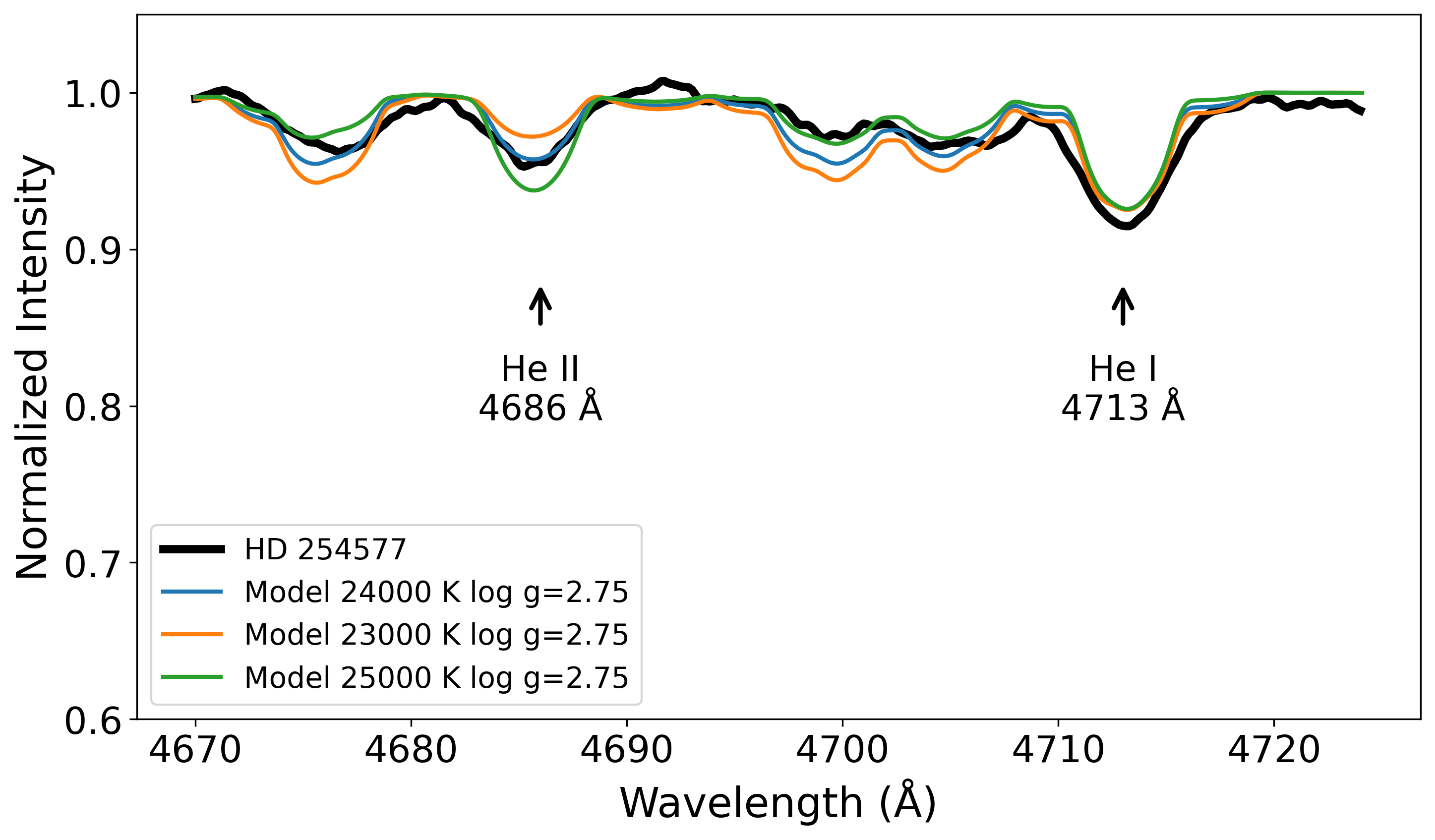}
\includegraphics[height=5.3 cm]{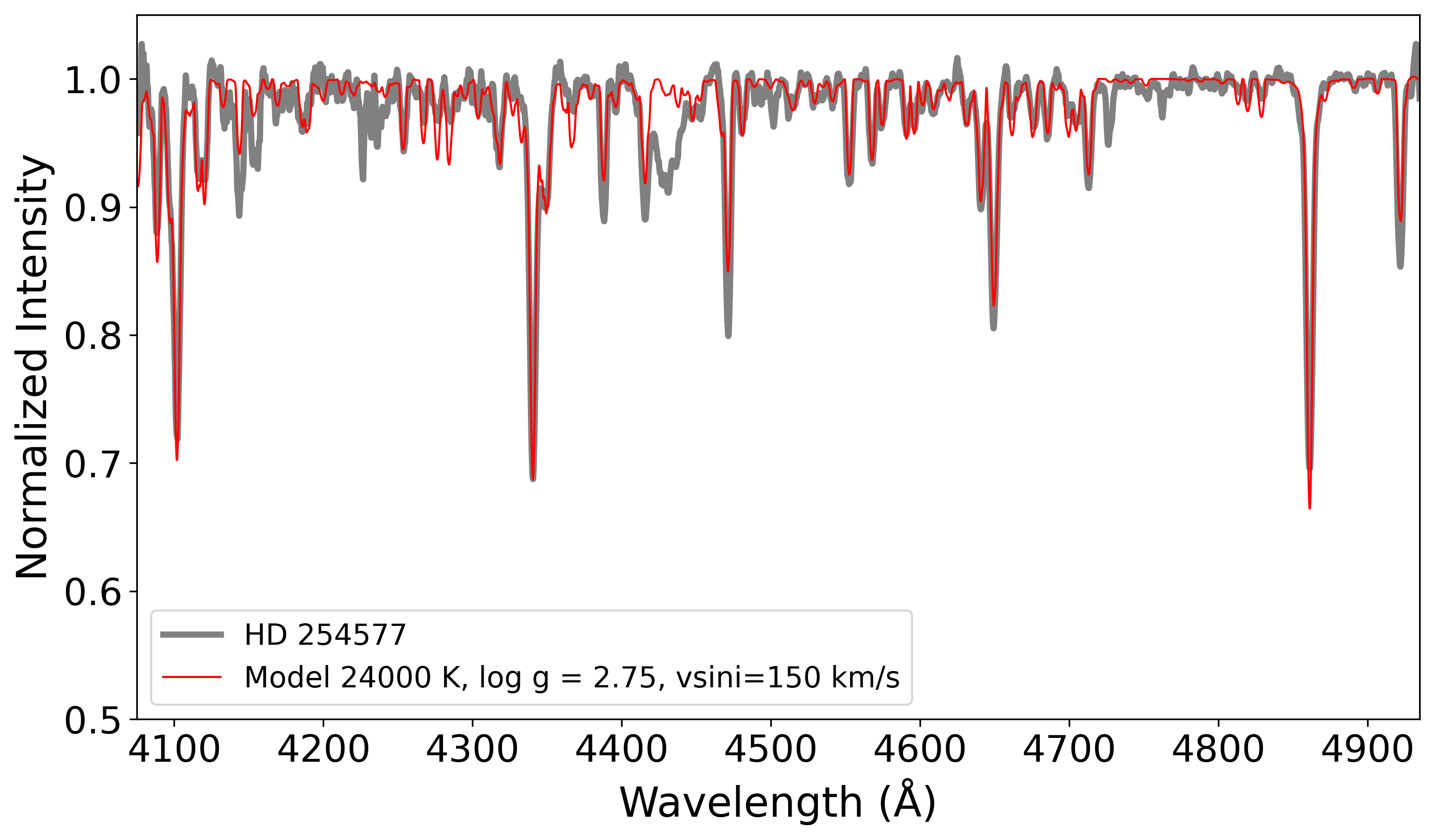}
\caption{\label{spec}Spectrum of HD~254577 with the overlayed model spectra of $T_\mathrm{eff}=23000$, 24000, and 25000 K with a fixed
surface gravity of $\log(g~\rm[cm/s^2])=2.75$, a projected rotation velocity of $v \sin i = 150$ km~s$^{-1}$ and a single micro-turbulence velocity of $\xi=10$ km~s$^{-1}$. 
\textit{Top Left:} The wavelength range of 
$4075-4140$ {\AA}.  Si~$\textsc{iv}$ $\lambda 4089,4116$ lines are prominent temperature indicators. 
The spectrum shows a consistency with the $T_\mathrm{eff}=24000$ K model at the Si~$\textsc{iv}$ $\lambda 4116$/ He~$\textsc{i}$ $\lambda 4121$ ratio.
\textit{Top Right:} 
The wavelength range of $4540-4660$ {\AA}. 
The spectrum shows a consistency again with the T$=24000$ K model at the Si~$\textsc{iv}$ $\lambda 4089$/ Si~$\textsc{iii}$ $\lambda 4553$ ratio.
\textit{Bottom Left:} 
The wavelength range of $4670-4725$ {\AA}. 
The spectrum shows a consistency again with the $T_\mathrm{eff}=24000$ K model at the He~$\textsc{ii}$ $\lambda 4686$/ He~$\textsc{i}$ $\lambda 4713$ ratio.
\textit{Bottom Right:} The spectrum of HD~254577 with thew  model spectrum of $T_\mathrm{eff}=24000$ K, $\log(g~\mathrm{[cm/s^2]})=2.75$, and $\xi=10$ km~s$^{-1}$ overlayed.
}
\end{figure*}


The spectra were compared with the non-local thermodynamic 
(NLTE), line-blanketed, plane-parallel synthetic spectra (Fig. \ref{spec})
which are calculated from the TLUSTY model atmospheres%
\footnote{http://tlusty.oca.eu/Tlusty2002/tlusty-frames-BS06.html} 
\citep{1995ApJ...439..875H}.
The models are available for a micro-turbulence velocity of 
$\xi=10$ and $\xi=2$ km~s$^{-1}$ and solar metallicity 
with a grid resolution of 1000 K in temperature and 0.25 dex for the surface gravity, $\log(g)$. The grid resolution sets the uncertainties of the measurements.

The model spectra were convolved with both the instrumental and rotational broadening functions.
We determined the instrumental line-spread function from the arc exposures, where the emission lines are well described by a Gaussian profile. 
For example, the full width at half maximum (FWHM) around H$\beta$ is 0.5 \AA ($\sim2.3$ pixels). 
Next, we applied a \citet{Gray2005} rotational profile, adopting a linear limb-darkening coefficient of 
$\epsilon = 0.6$ for rotational velocities between 50 and 200 km~s$^{-1}$ in steps of 10 km~s$^{-1}$.
This two-step convolution procedure ensures that the synthetic spectra reproduce the correct core–wing morphology of the hydrogen Balmer lines, whose wings are dominated by Stark broadening and are strongly sensitive to surface gravity.

We determined the surface gravity from the H$\beta$ wings by visual comparison, and effective temperature from the silicon (Si~$\textsc{iv/iii}$) ionization balance, with helium (He~$\textsc{ii/i}$) used as a secondary check. Estimates from the two ratios generally agree to within $\pm$1000 K.

We measured the RV of the runaway star in different epochs.
The observed spectrum with the highest S/N ratio, taken on 2015, Mar 18, 
were applied Fourier cross-correlation with template model spectrum using the IRAF task \textit{fxcor} after being shifted in the wavelength space for the heliocentric RV.
The wavelength ranges used for cross-correlation are 
$4295-4928$, $5865-5885$, $6540-6590$, and $6650-6715$ {\AA}.
We chose these wavelength ranges considering the data quality 
and the existence of prominent spectral features inside.
We measured the RVs from other spectra of the runaway star with the same method,
yet using the 2015, Mar 18, spectrum as a template to maintain higher correlation heights (Table \ref{rav}). 

\subsection{SNR Spectra}
We also present the results of the optical emission associated 
with SNR G$189.6+03.3$ and IC 443 based on LAMOST low resolution spectra. 
We used all available spectra of SNR G$189.6+03.3$ and IC 443. 
Among these, we chose 94 stellar spectra from 85 locations with a low continuum level ($S/N<10$). 
Spectra were taken in 2017, and each DIT is 900 seconds.

\begin{figure*}
\centering
\includegraphics[width=\textwidth]{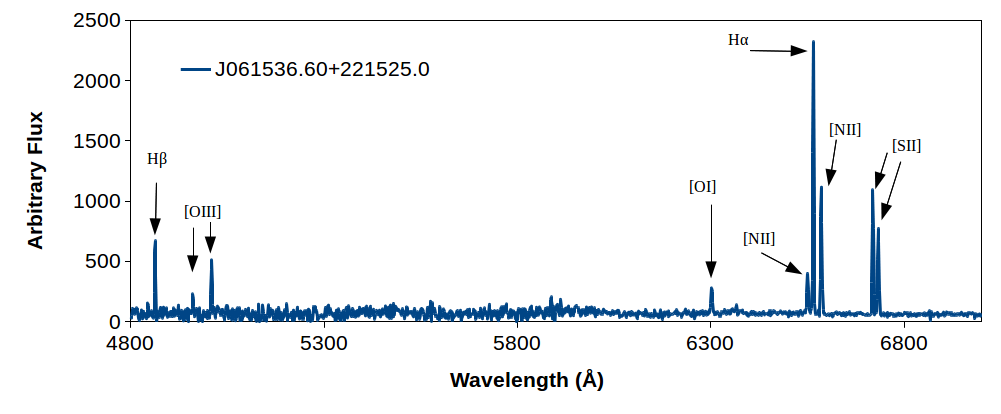}
\caption{\label{snrspec}  LAMOST Spectrum of the pointing at  $\mathrm{\alpha=93.902500^\circ}$ $\mathrm{\delta=+22.256944^\circ}$. The prominent emission lines are H$\beta$ $\lambda$4861, H$\alpha$ $\lambda$6563, and forbidden lines [O~{\sc iii}] $\lambda\lambda$4959/5007, 
[O~{\sc i}] $\lambda \lambda$6300,
[N~{\sc ii}] $\lambda \lambda$6548/6584, 
and [S~{\sc ii}] $\lambda \lambda$6716/6731.
}
\end{figure*}

SNRs show a typical nebular spectrum dominated by strong emission lines of H and forbidden transitions of metals such as O, N, and S (Fig. \ref{snrspec}).
A wavelength range of $4200-6900$ {\AA} is sufficient to detect important features
and calculate optical parameters from their fluxes \citep[see][]{Fe2020, Bo2022, Ba2023, Pa2024}.
LAMOST spectra show hydrogen Balmer lines, H$\beta$ $\lambda$4861, H$\alpha$ $\lambda$6563{\AA}, and forbidden lines [O~{\sc iii}] $\lambda \lambda$4959/5007, 
[N~{\sc ii}] $\lambda \lambda$6548/6584, 
and [S~{\sc ii}] $\lambda \lambda$6716/6731 (Table \ref{Lines}). 

We measured each flux individually from the already reduced spectra of the LAMOST data using IRAF. Since they are reduced data, no flux calibration was performed. The fluxes of H$\alpha$ $\lambda$6563 are taken as 100. We normalized the fluxes of the spectra to H$\alpha$ $\lambda$6563 and calculated the optical parameters of the SNRs (Table \ref{parameters}).

For an accuracy check, we calculated the parameters of a single pointing, 
J061524.40+223308.8 from 10 different LAMOST spectra, 
and we found that the standard deviation of 
$[$S$~${\sc ii}$]$/H$\alpha$ ratios is 0.05 where the average ratio is 1.02.
Yet the standard deviation of 
$[$O$~${\sc iii}$]$/H$\beta$ ratio is 1.27 where the average ratio is 4.03.
The low quality of $[$O$~${\sc iii}$]$/H$\beta$ ratio is due to the lower signal-to-noise ratio in the blue part of the spectrum and the weaker nature of these lines in the nebular spectrum (Fig. \ref{snrspec}).  
Nevertheless, these spectra provide us with highly accurate 
$[$S$~${\sc ii}$]$/H$\alpha$ and $[$S$~${\sc ii}$]$ 6716/6731 ratios.
\begin{figure*}
\centering
\includegraphics[width=0.47\textwidth]{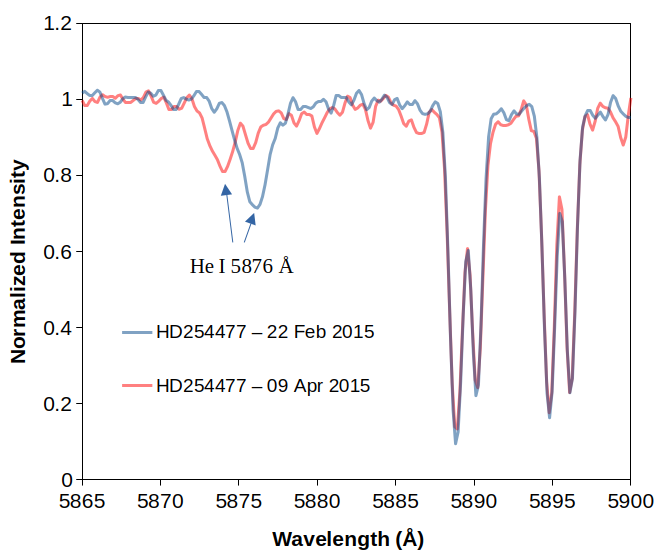}
\includegraphics[width=0.51\textwidth]{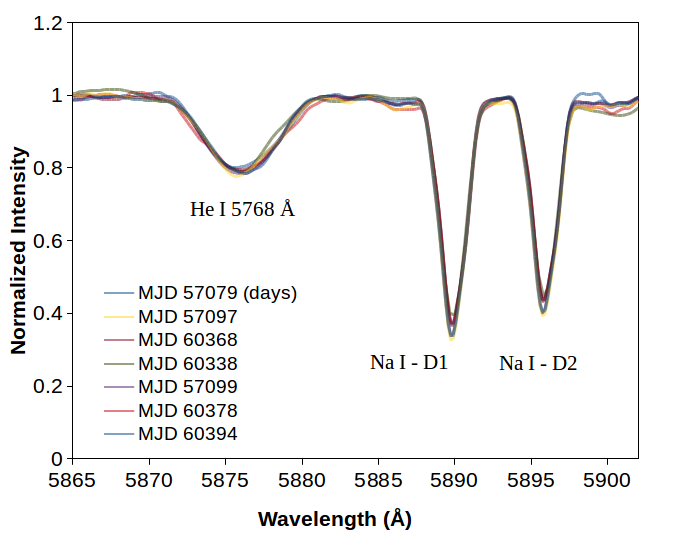}
\caption{\label{477rv}\textit{Left:} FLECHAS spectra of HD254477. Whereas the interstellar medium Na$~\textsc{i}$ lines show no variation 
between epochs, the Doppler variation of He$~\textsc{i}$ $\lambda5876$ line is clear. \textit{Right:} FLECHAS spectra of HD254577. He$~\textsc{i}$ $\lambda5876$ triplet does not show a significant variation as well as the Na$~\textsc{i}$ lines.}
\end{figure*}
\section{Results and Discussion }

Based on the comparison with the synthetic TLUSTY spectra, we determined that HD~254577 is a hot, evolved star with a $T_\mathrm{eff}=24000\pm1000$ K, and
$\log(g~\rm[cm/s^2])=2.75\pm0.25$ for solar metallicity and with an adopted microturbulence velocity of $\xi=10$ km~s$^{-1}$ (Fig. \ref{spec}).
The projected rotational velocity of the star is $v \sin i = 150\pm20$ km~s$^{-1}$.
The weighted mean heliocentric RV of HD~254577 is $RV_\mathrm{helio}=17.1\pm2.4$ km~s$^{-1}$ with a standard deviation of $3.9$ km~s$^{-1}$.
Unfortunately, no RV measurement is present for the cluster; therefore, we do not have a reference RV.
Thus, we calculated the peculiar RV using a flat Galactic rotation curve for a solar distance of 8.5 kpc and an orbital velocity of 220 km~s$^{-1}$. 
We found $RV_\mathrm{pec}=-0.8\pm3.9$ km~s$^{-1}$. 
The peculiar 3-D velocity is then $v_\mathrm{pec}=31.3^{+1.2}_{-0.9}$ km~s$^{-1}$,
above the typical runaway star velocity criterion, 30 km~s$^{-1}$, yet, lower than $\sim40$ km~s$^{-1}$ threshold
suggested by \cite{2001ApJ...555..364B} and \cite{2024ApJS..272...45G}.
Nevertheless, according to \cite{2019A&A...624A..66R}, the SN ejected massive runaway stars can have even lower velocities.
HD~254477, the possible cluster member, shows a clear RV variation of 
$\sim60$ km~s$^{-1}$ suggesting a spectroscopic binary nature.
In contrast, the runaway star does not show an RV variation outside the measurement uncertainty (Table \ref{rav}), 
implying that it is a single star, as expected primarily for a runaway star
(Fig. \ref{477rv}).  
\begin{table*}
\centering
\caption{\label{sp}%
Measured parameters of the observed stars with LAMOST. The spectral types suggested for the stars in this work (SpT\_tw), with the most precise determination available in the literature (SpT\_lit), are given. References: (1) \cite{1955ApJ...121...24C} and (2) \cite{2002A&A...386..709F}
}
\begin{tabular}{l l l l l}
\toprule
\textit{Gaia} DR3
& $T_{\rm{eff}}$ (K)  
& $\log(g)$ 
& SpT\_tw
& SpT\_lit     \\
\midrule
3377027132948376704  &$13500\pm500 $ &$3.75\pm0.25$  &B$7\pm1$IV            & -            \\
3377026961149682432  &$16000\pm1000$ &$4.25\pm0.25$  &B$4\pm1$V             & -            \\
3377004215002923904  &$19000\pm1000$ &$4.25\pm0.25$  &B$2.5\pm0.5$V         & -            \\
3377002840613390592  &$11500\pm500 $ &$4.00\pm0.25$  &B$9\pm1$V             & -            \\
3376999709578280064  &$25000\pm2000$ &$4.25\pm0.25$  &B$0.5\pm0.5$V         & B2/3 III (1) \\
3377029950446864896  &$15000\pm1000$ &$4.25\pm0.25$  &B$5\pm1$V             & -            \\
3377054620738994816  &$10000\pm500 $ &$3.00\pm0.25$  &B$9.5\pm0.5$III       & -            \\
3377012731919116416  &$26000\pm2000$ &$4.25\pm0.25$  &B$0.5\pm0.5$V         & B8 (2)       \\
3377012697559377664  &$15000\pm1000$ &$4.25\pm0.25$  &B$5\pm1$V             & -            \\
3377049737357365760  &$15000\pm1000$ &$4.25\pm0.25$  &B$5\pm1$IV            & -            \\
3377037230412601472  &$10000\pm500 $ &$2.75\pm0.25$  &B$9.5\pm0.5$II        & -            \\
3377039433734571520  &$15000\pm1000$ &$4.25\pm0.25$  &B$5\pm1$IV            & -            \\
3377044690774512896  &$20000\pm2000$ &$4.25\pm0.25$  &B$2\pm1$V             & -            \\
3377046168243238528  &$13000\pm500 $ &$4.25\pm0.25$  &B$7\pm1$V             & -            \\
3377033936176712192  &$12000\pm500 $ &$3.50\pm0.25$  &B$8\pm1$IV            & -            \\
3377040464526993920  &$19000\pm1000$ &$4.25\pm0.25$  &B$2.5\pm0.5$V         & -            \\
3376852070081305344  &$10000\pm500 $ &$4.00\pm0.25$  &B$9.5\pm0.5$IV        & -            \\
\bottomrule
\end{tabular}
\end{table*}

\subsection{Parent Cluster}
\begin{figure*}
\centering
\includegraphics[height=6.4 cm]{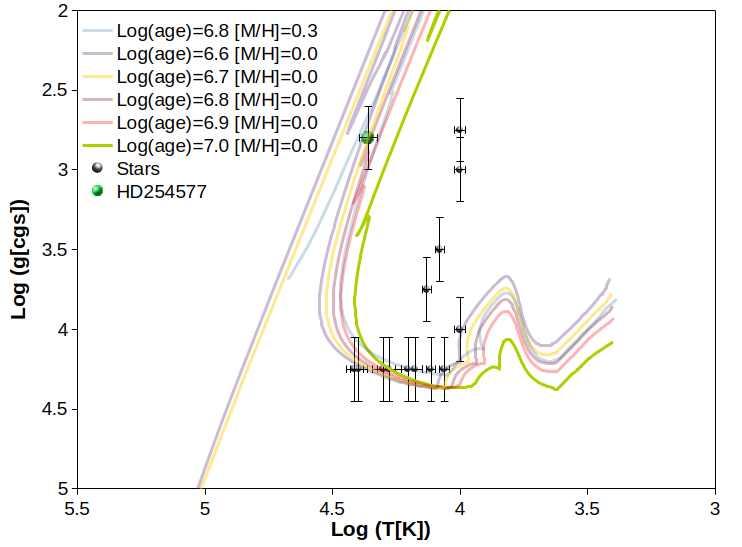}
\includegraphics[height=6.55 cm]{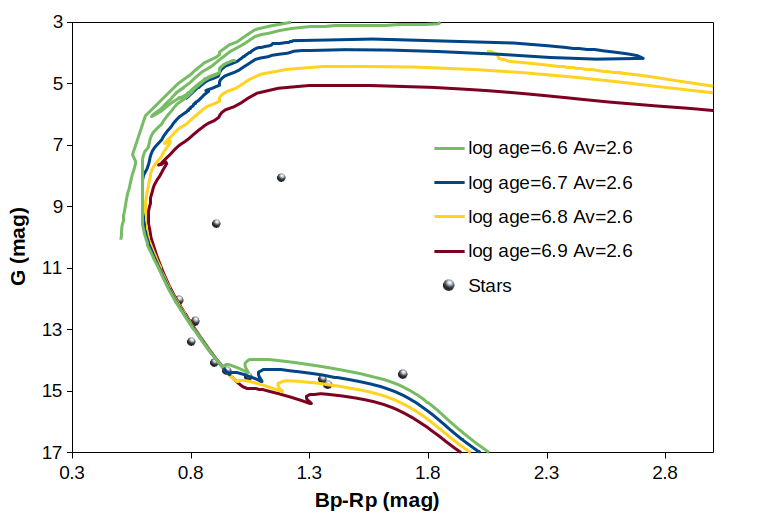}
\caption{\label{parent}\textit{Left:} $T_\mathrm{eff}$ vs. $\log(g)$ diagram of the runaway star and OB-type stars within 10 pc from the runaway star. \textit{Right:} The color-magnitude diagram of the runaway star and the possible cluster members. Isochrones are set for a 1700 pc distance. The bright stars out of the isochrones are the runaway star and HD~254477. 
}
\end{figure*}

HD~254577 most likely shares the same origin with the stars inside the small circle in Fig. \ref{f_g189} and some of the stars within 10 pc. 
Based on the archival LAMOST spectra, we determined the atmospheric parameters of 
17 OB-type stars within 10 pc from the runaway star (Table \ref{sp}). 
Two of the stars, HD~254477 ($T_\mathrm{eff}=26000$ K, $\log(g~\rm[cm/s^2])=4.25$) and J061731.40+222555.2 ($T_\mathrm{eff}=16000$ K, $\log(g~\rm[cm/s^2])=4.25$), are inside the restricted region of the possible cluster. 
Hence, we set the color-magnitude diagram (CMD), that is G vs. G$_{BP}-$G$_{RP}$ 
of these stars, and the runaway star, and compared it with theoretical isochrones.
We produced isochrones for solar metallicity with the CMD 3.7%
\footnote{http://stev.oapd.inaf.it/cgi-bin/cmd\_3.7}
website using the "Parsec 1.2S" stellar evolutionary tracks
\citep{2015MNRAS.452.1068C} with bolometric corrections from \cite{2008PASP..120..583G}   
and the extinction curve from  \cite{1989ApJ...345..245C} and \cite{1994ApJ...422..158O} with an extinction law R$_V=3.1$.
We used the canonical two-part power-law initial mass function introduced in \cite{2001MNRAS.322..231K, 2002Sci...295...82K},
to compute the stellar occupation and masses along the isochrones.
Apart from HD~254477 and HD~254577, the cluster stars are mainly aligned on the CMD (Fig. \ref{parent}).
However, the position of the two bright stars implies an age of log(age[yr])=7.8 
where their observed and theoretical atmospheric parameters do not match.
This is because the spatial variation of the extinction and the total-to-selective absorption ratio is quick and decisive \citep{2013ApJ...768L...8K}.
Therefore, we determined the cluster parameters by fitting isochrones on $T_\mathrm{eff}$ vs. $\log(g)$ diagram.
Although surface gravities have large uncertainties, temperatures were determined precisely enough to restrict the cluster parameters,
and we found that the cluster age is log(age[yr])$=6.7\pm0.1$ (Fig. \ref{parent}).
We identified the progenitor as the most evolved, and thus the most massive, star at the terminal point of each isochrone.
The mass of the runaway star varies from 26 to 41 M$_\odot$, 
while its zero-age main sequence mass (ZAMS) is between 28 and 54 M$_\odot$.   
The progenitor has a ZAMS mass of 31-64 M$_\odot$, and a final mass of 16-21 M$_\odot$ considering a mass loss for solar metallicity (Table \ref{mass}).
\begin{table}                                                             
\centering
\caption{\label{mass}The initial and current (final) masses for the runaway star and the progenitor are given for different cluster ages. The masses are in M$_\odot$.}
\begin{tabular}{c c c c c}
\toprule
$\log(\text{age[yr]})$	
&$M_\mathrm{ini-run}$
&$M_\mathrm{ini-pro}$
&$M_\mathrm{run}$
&$M_\mathrm{pro}$ \\
\midrule
6.6	&$46\pm1$ \& $54$	&64	&41 \& 32	&21 \\
6.7	&35	&42	&32	&16\\
6.8	&28	&31	&26	&20\\
\bottomrule
\end{tabular}
\end{table}

A massive progenitor ($>30$ M$_\odot$) we propose in this work confirms \cite{2018A&A...615A.157G} who explained the jet-like structures of IC~443 
with the SN of a massive progenitor having a mass of $\sim30$ M$_\odot$. 
It might also explain the high abundance of $\alpha-$process elements, detected in \cite{2008A&A...485..777T}.

\subsection{Kinematics}
\label{Kinematics}
\begin{table}
\centering
\caption{\label{psrvel}Parameters found by tracing back the runaway star.
The coordinates of the explosion center and the related NS transverse velocities are given for various SNR-NS ages.}
\begin{tabular}{c l l c}
\toprule
Age 
&Exp. R.A.  & Exp. Dec. &$v_\mathrm{NS}$   \\
(kyr) &(deg) &(deg) &(km~s$^{-1}$)\\
\midrule
4	&94.47196$\pm$0.00012	&22.40876$\pm$0.00029 &$1391.0^{+2.7}_{-1.6}$\\[3pt]
10	&94.46499$\pm$0.00029	&22.40818$\pm$0.00073	&$537.9^{+1.9}_{-0.7}$\\[3pt]
15	&94.45918$\pm$0.00044	&22.40769$\pm$0.00110	&$348.4^{+1.7}_{-0.5}$\\[3pt]
20	&94.45337$\pm$0.00058	&22.40721$\pm$0.00146	&$253.6^{+1.6}_{-0.4}$\\[3pt]
30	&94.44176$\pm$0.00087	&22.40624$\pm$0.00219	&$158.8^{+1.6}_{-0.3}$\\
\bottomrule
\end{tabular}
\end{table}

\begin{figure}
\includegraphics[width=\columnwidth]{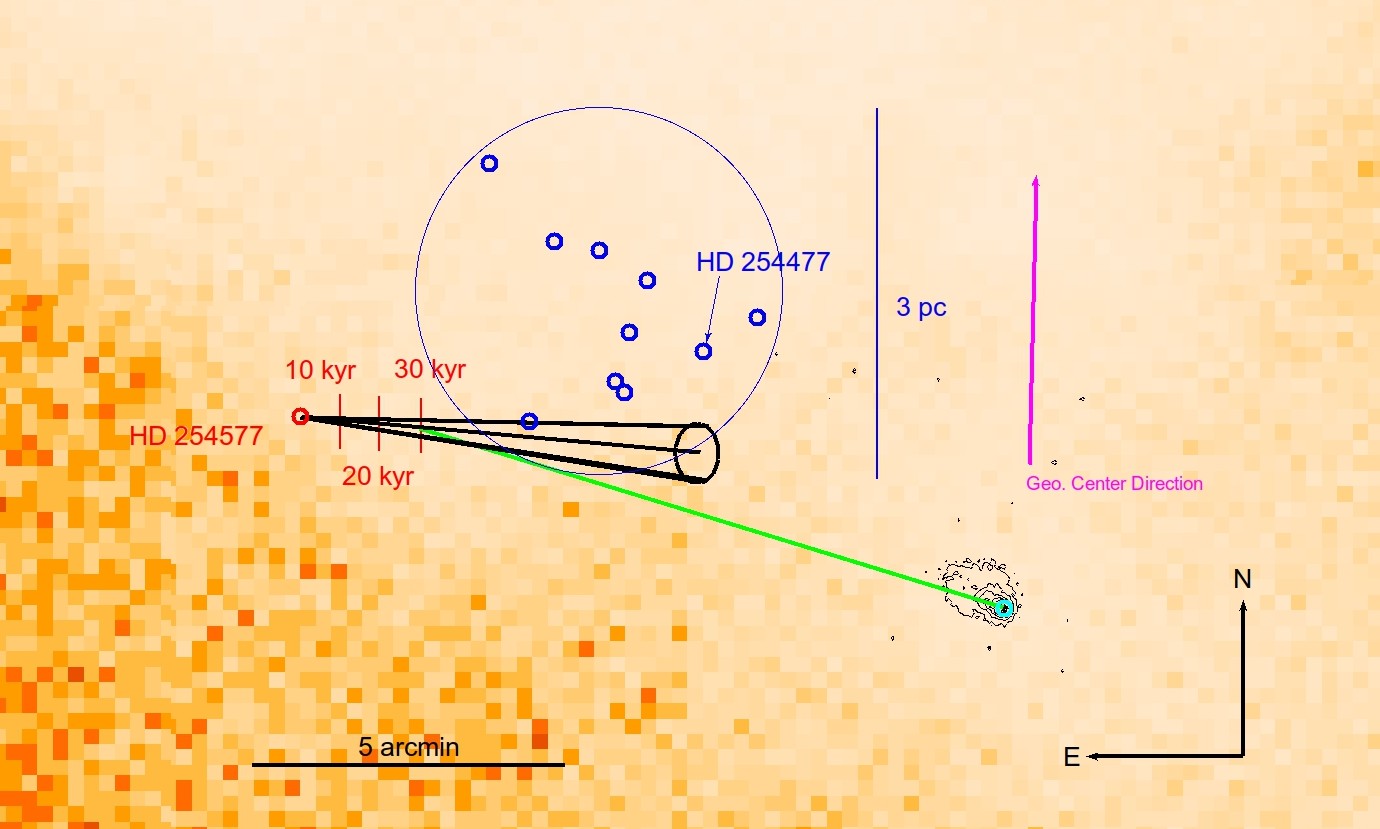}
\caption{\label{kin}Tracing back of the proper motion of HD~254577 w.r.t the cluster motion (black cone)
sketched on the ROSAT PSPC (0.1--2.4 keV) image. The explosion sites for 10, 20 and 30 kyr are shown in red lines. The big blue circle shows the extension of the possible open cluster of which member are denoted by small blue circles.
The Chandra ACIS contours of the NS (cyan circle) are overlayed in black.
The green line roughly shows the direction of the NS motion from the 30 kyr explosion site, while the magenta vector shows the direction of the geometrical center of IC~443.}
\end{figure}

We traced back the peculiar proper motion of the runaway star
to find the explosion location and then calculated the transverse velocity of the NS for different possible ages.
We multiplied the peculiar proper motion by five different SNR ages between $4$ and $30$ kyr, and subtracted the angular distance found from the position of the star (Fig. \ref{kin}).
The NS coordinates, $\alpha=94.27190^\circ$ and $\delta=+22.35818^\circ$ with a positional error of 0.6 arcsec, are retrieved from \cite{2012ApJ...756...27L}.
We measured the angular separation of the NS from each explosion location found.
As these positions are already dependent on the time passed since the explosion, 
we calculated the hypothetical proper motion of the NS
by simply dividing the separation by the corresponding age.
Finally, we calculated the transverse velocities in km~s$^{-1}$ for a fixed distance of 1.7 kpc.
The NS velocities between $10-20$ kyr ($v_\mathrm{trans}=254-538$ km~s$^{-1}$) (Table \ref{psrvel}) are consistent with typical pulsar velocities.
The predicted velocity of the NS, 230 km~s$^{-1}$ \citep{2006ApJ...648.1037G}, 
can be reached at higher ages. 
For a possible earlier age of $4$ kyr \citep{2008A&A...485..777T}, 
the NS is moving exceptionally fast, with a velocity of $\sim1400$ km~s$^{-1}$.
Considering the age estimates of the SNR in the literature and the NS velocity distribution, we suggest that the highly probable age of the SNR is $10-20$ kyr.

The fact that the NS can attain typical space velocities for reasonable SNR ages provides further support for the pre-SN binary scenario we propose.
On the other hand, although the runaway star is moving away from the cluster, we note that the past position of the star is not located inside the possible cluster circumference within $30$ kyr.
For larger ages, the NS velocity drops below 159 km~s$^{-1}$.
Yet, the iron core-collapse origin of the SNR is strongly suggested, 
and such an SN generally forms a high NS-kick.
The pre-SN system can still be genetically connected to the cluster, 
but slightly drifted away owing to the cluster dynamics, 
or simply the cluster is loosely bound, and its radius is more extensive.

\subsection{Interstellar medium absorption lines}
\begin{figure*}
\centering
\includegraphics[width=0.33\textwidth]{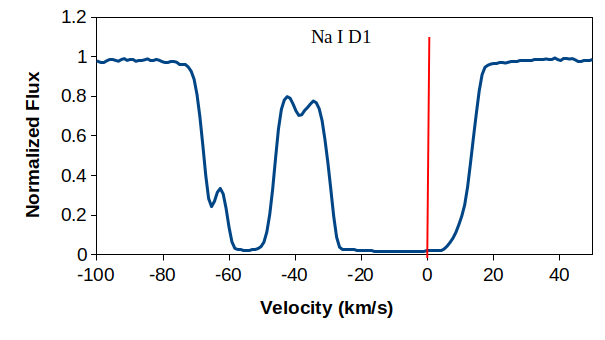}
\includegraphics[width=0.33\textwidth]{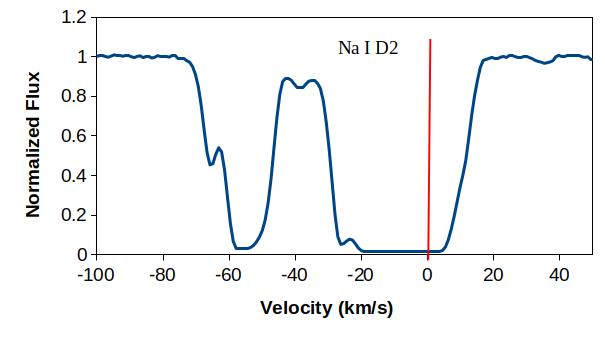}
\includegraphics[width=0.33\textwidth]{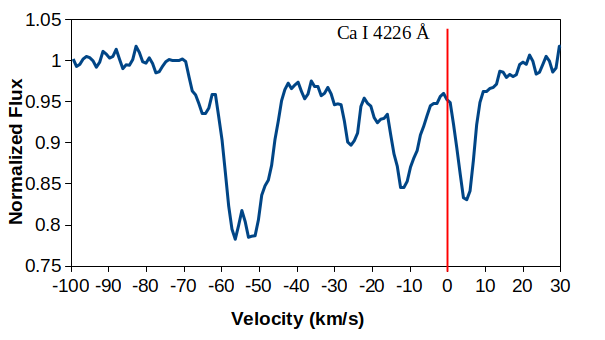}\\
\includegraphics[width=0.33\textwidth]{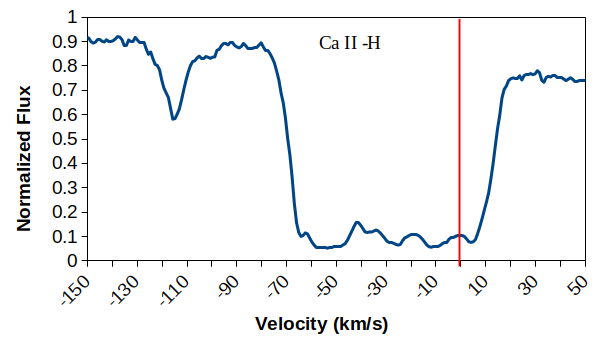}
\includegraphics[width=0.33\textwidth]{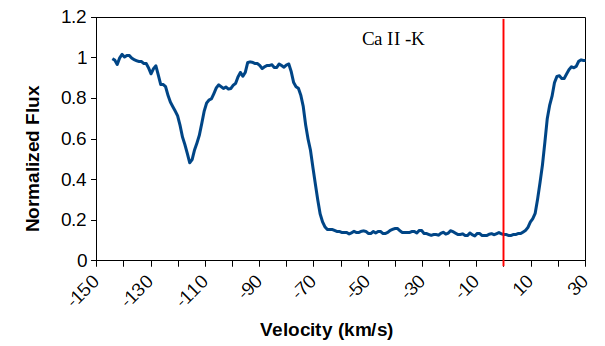}
\caption{\label{ismlines}ISM lines towards HD~254577 detected in the HDS spectrum.}
\end{figure*}

The high-velocity interstellar medium (ISM) absorption lines on the stellar spectra towards IC~443 were extensively studied in 
\cite{2003A&A...408..545W, 2009ApJ...696.1533H, 2012ApJ...750L..15T, 2020ApJ...897...83R}.
Several stars behind the SNR show blue- and red-shifted ISM absorption lines on their spectrum with velocities up to $\sim$100 km~s$^{-1}$.
Although we did not find a new high-velocity absorption feature, we studied the prominent ISM features in the HDS spectrum of HD~254577.
Assuming the observed features are dominated by the molecular cloud interacting with the SNR, we calculated the rest velocity for solar peculiar motion and the galactic rotation as described in the section \ref{Kinematics}.
We found a characteristic velocity of 17.9 km~s$^{-1}$ and displayed this reference 
as the zero velocity in Fig. \ref{ismlines}.
Therefore, the largest positive velocity of ISM lines is around 10 km~s$^{-1}$, which should not be considered as SNR acceleration.
On the other hand, there are blue-shifted features at $\sim -100$ km~s$^{-1}$ that
is most probably due to the SNR expansion.
This suggests that the runaway star is inside the SNR, so we do not detect any red-shifting ISM lines. This is another strong evidence that the runaway star is the pre-SN binary companion of the progenitor of IC~443.
Furthermore, the velocities found in \cite{2020ApJ...897...83R} for HD~254477, the possible cluster member, are highly similar to the runaway star, strengthening the idea that they are genetically connected.

\subsection{The SNR Parameters}
\begin{figure*}
\centering
\includegraphics[width=0.49\textwidth]{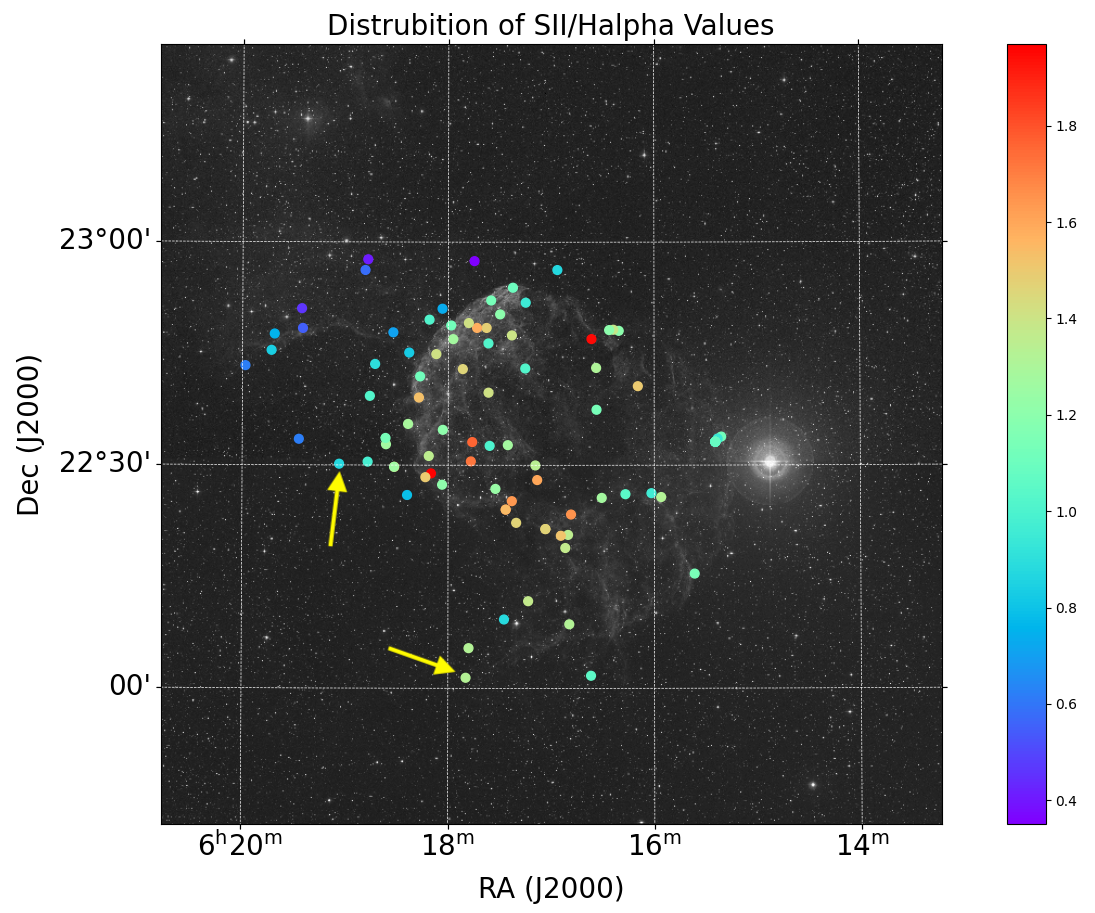}
\includegraphics[width=0.49\textwidth]{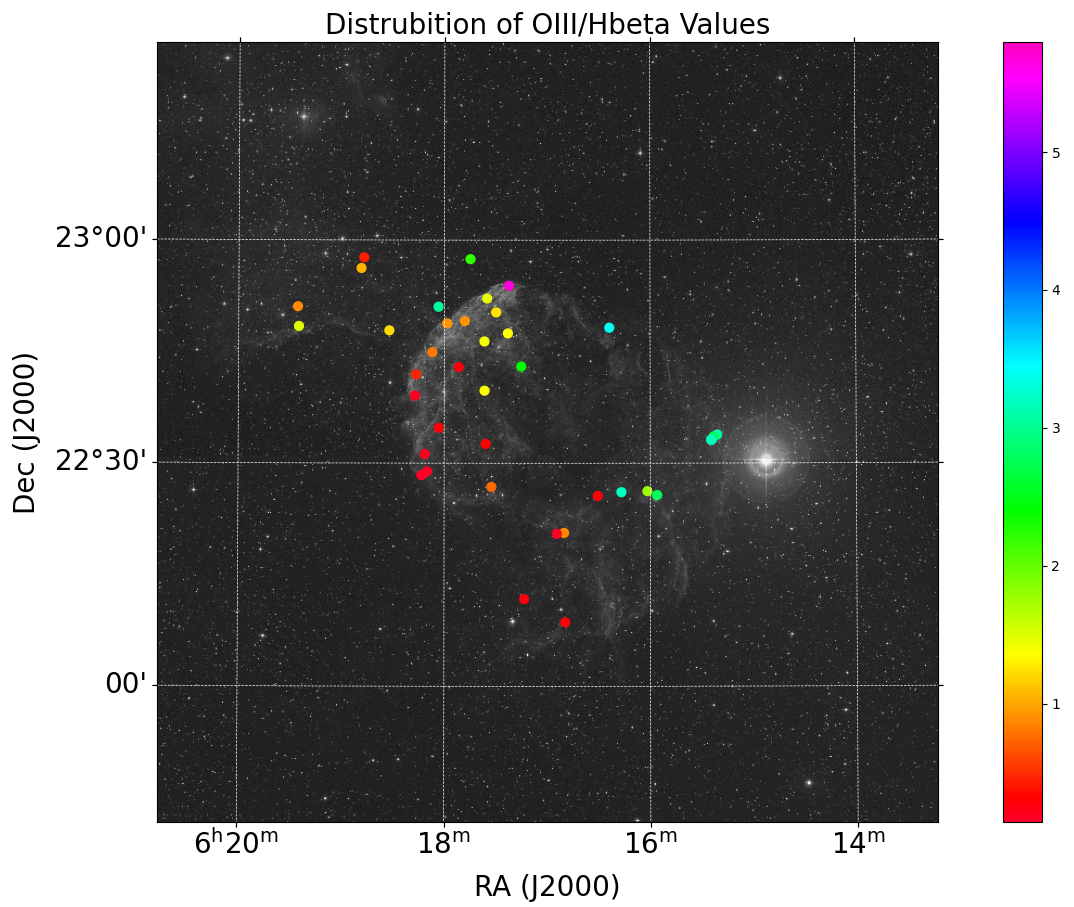}
\caption{\label{sii/ha}\textit{Left:} Measured $[$S$~${\sc ii}$]$/H$~\alpha$ ratios represented by the color bar. The region includes SNRs IC~443 and the western part of G$189.6+03.3$. Two spectrum positions indicated by yellow arrows are $\mathrm{\alpha=06h19m27.00s}$
$\mathrm{\delta=+22d33'30.0''}$, and $\mathrm{\alpha=06h17m49.70s}$
$\mathrm{\delta=+22d01'25.3''}$. 
The strong  $[$S$~${\sc ii}$]$ emission suggests
that these faint filaments belong to IC~443, not G$189.6+03.3$.
\textit{Right:} Same figure for $[$O$~${\sc iii}$]$/H$\beta$ ratios.
The $[$O$~${\sc iii}$]$/H$\beta$ ratios are generally higher in the west and in the north 
compared to the east and south.}
\end{figure*}

 We determined the physical parameters and the actual size of the SNR based on the Balmer, H$\beta$ $\lambda$4861, H$\alpha$ $\lambda$6563, and forbidden lines [O\,{\sc iii}] $\lambda \lambda$4959/5007, 
as well as [S\,{\sc ii}] $\lambda \lambda$6716/6731 emission lines 
(Table \ref{Lines}) in the LAMOST spectra taken from various pointings throughout the SNR.

For the northeast (NE) region, together with one additional pointing from the east (E) far outside of IC~443, we found that the ratio of $[\mathrm{S}~\textsc{ii}]/\mathrm{H}\alpha=0.61\pm0.14$, considerably smaller than in the main body of IC~443, which is $[\mathrm{S}~\textsc{ii}]/\mathrm{H}\alpha=1.25\pm0.38$, and is consistent with emission from an H~{\sc ii} region and possibly the older SNR G$189.6+03.3$ (Table (\ref{parameters})). For the region slightly outside of IC~443, we measured $[\mathrm{S}~\textsc{ii}]/\mathrm{H}\alpha$ ratios higher than $0.8$ from the long-slit spectra, which supports that the emission originates from shock-heated gas \citep{Fe85}. 
We also detected ratios as high as 1.33 outside the bright filaments of IC~443, indicating that the actual size might be larger (Fig. \ref{sii/ha}).

The $[O~\textsc{iii}]$ $\lambda$5007 line relative to H$\beta$ is mainly 
an indicator of the mean level of ionization and temperature \citep{Do84}. Based on a planar shock model of \citet{Har87}, we calculated the $[\mathrm{O}~\textsc{iii}]/\mathrm{H}\beta$ ratios. For the NE region, the observed ratio of $[\mathrm{O}~\textsc{iii}]/\mathrm{H}\beta=1.19\pm0.42$ indicates a shock velocity of 72 km~s$^{-1}$ with a complete recombination zone (see \citealt{Ra79, Sh79}). 
The weak $[\mathrm{O}~\textsc{iii}]$ emission may be explained by slow shocks propagating into the ISM \citep{Ra88}. 
For the whole region, the observed ratio of 
$[\mathrm{O}~\textsc{iii}]/\mathrm{H}\beta=2.81\pm2.26$  
that shows the relatively low $T$ and a slow expansion of shock \citep{OsFe06}. 
Due to the low temperature, slow shock velocity is observed in the SE region, while relatively fast shock propagation continues in the W and N regions.

We measured the electron density ($N_{\rm e}$) based on $[\mathrm{S}~\textsc{ii}]$ $6717/6731$ flux ratios \citep{OsFe06} using the \texttt{temden} task of the \texttt{nebular} package \citep{1995PASP..107..896S}, 
assuming an electron temperature ($T$) of $10^{4}$ K. 
For the NE and E regions outside IC~443, 
we found an average electron density of 224$\pm75$ cm$^{-3}$.
On the other hand, in the IC~443 part, 
the average density is 146$\pm92$ cm$^{-3}$. 
$N_{\rm e}$ varies between 4 to 1257 cm$^{-3}$ throughout the whole region, 
indicating the ionized gas heated by shock.

We derived a logarithmic extinction $c$ for the NE and one additional pointing from the E regions 
with observed H$\alpha$/H$\beta$ flux ratio. 
We measured a color excess of $E(B-V)
=0.47\pm0.11$ and 0.92$\pm0.39$ for the NE and E regions, respectively. 
Using the relation $N_{\rm H} = 5.4 \times$ $10^{21}$ $\times$ $E(B-V)$ \citep{Pr95}, 
we calculated a total column density of $\sim$4.97$\times$ $10^{21}$ cm$^{-2}$ for the E region while, the total column density is $\sim$2.54$\times$ $10^{21}$ cm$^{-2}$ in the NE region.
This implies that the H~{\sc ii} regions in the NE and E regions 
are much farther away than the SNR, 
and optical emission from these regions may not be related to IC~443.
The values found in this section also coincide with the results of \cite{ba2024}.

\begin{figure}
\includegraphics[width=\columnwidth]{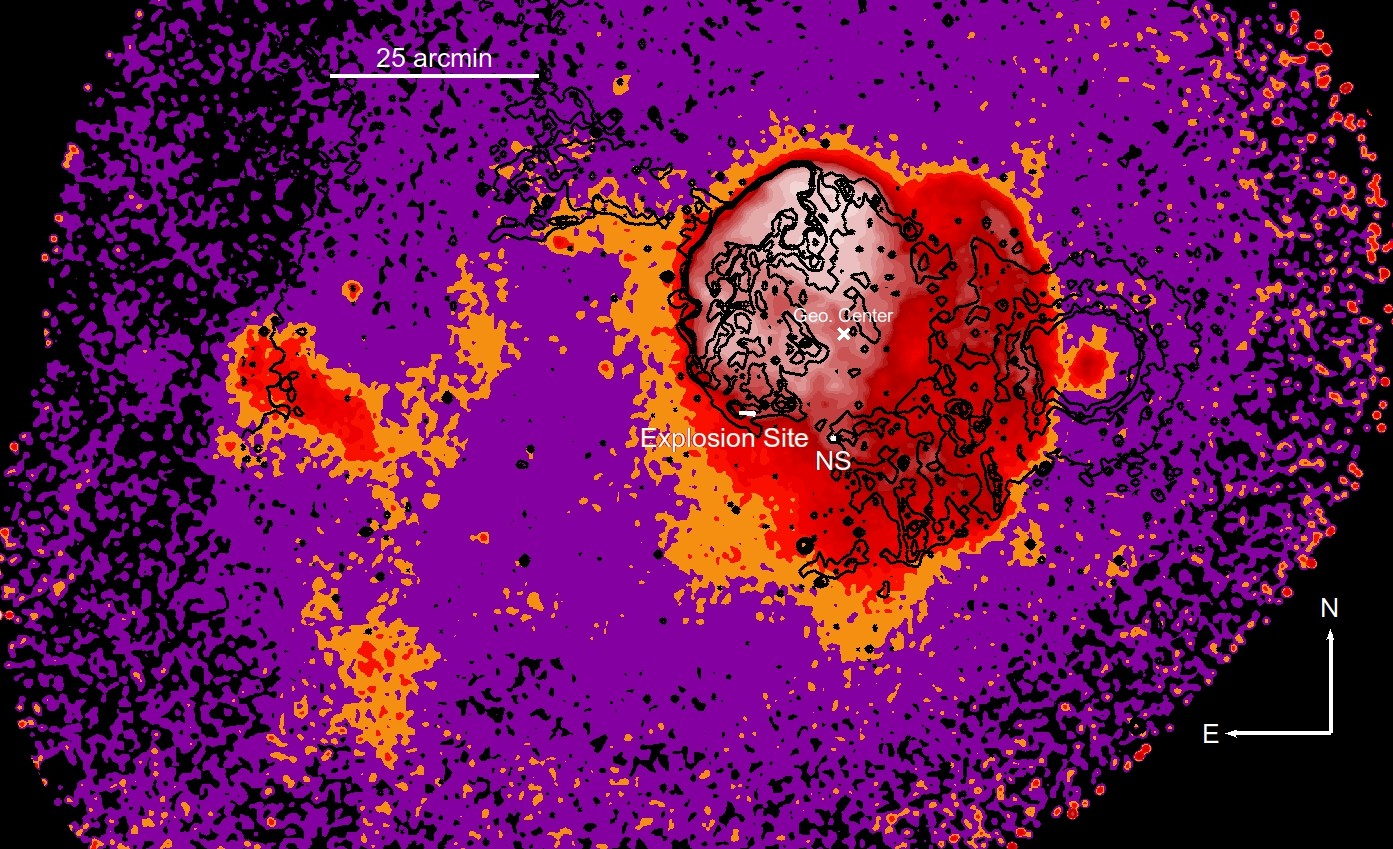}
\caption{\label{ic443}ROSAT PSPC (0.1--2.4 keV) image of IC~443 region. 
The black contours overlayed is the bright optical filaments.
The explosion sites from 0 to 30 kyr are shown as a white cone.
The geometrical center is denoted by a white cross and the NS by a white circle.
The SNR is in a complex region together with SNR G$189.6+03.3$ and H$~\textsc{ii}$
regions. However, the true extension in the east-southeast is probably significantly large and the explosion sites we found are closer to the real geometrical center.
}
\end{figure}

The extended geometry of IC~443 is also noticeable in X-rays (Fig. \ref{ic443}). 
Although the SNR entirely overlaps with G$189.6+03.3$, the high $[\mathrm{S}~\textsc{ii}]/\mathrm{H}\alpha$ extension (Fig. \ref{sii/ha}) coincides with the X-ray contours gradually decreasing in brightness toward the east. 
The position of the explosion is highly separated from the geometrical center of the
bright IC~443. 
However, \cite{2021A&A...649A..14U} explains how IC~443 expands asymmetrically 
due to the interaction with dense and inhomogeneously distributed molecular clouds and assumes that the explosion center is not the geometrical center, but the NS.

The shock wave of the SNR has either lost its energy or is expanding freely toward the east, making such an asymmetric shape. The lower values of $[\mathrm{O}~\textsc{iii}]/\mathrm{H}\beta$ ratios in the east and south may  
indicate that the shock has lost a significant amount of energy and is expanding slowly in this direction due to the interaction with the dense molecular cloud.

\subsection{Pre-SN Binary}
\begin{table}
\centering
\caption{\label{preSN}Calculated parameters of the pre-SN binary for the peculiar velocity of 31 km~s$^{-1}$.
The binary separation, the period, the orbital velocity of the primary, and the Roche-Lobe radius for different configurations are given, respectively.}
\begin{tabular}{@{}c c c c c@{}}
\toprule
Conf. (M$_\odot$) &Sep. (R$_\odot$)   & P (days) &$v_\mathrm{orb}$ (km~s$^{-1}$) & RL (R$_\odot$) \\
\midrule
41 \& 5 &110	&20	    &254	&24 \\
41 \& 10 &440	&141	&127	&117 \\
41 \& 15 &989	&433	&85	&293\\
41 \& 21 &1939	&1073	&61	&626\\
32 \& 5 &141	&31	    &198	&33\\
32 \& 10 &563	&219	&99	&160\\
32 \& 16 &1442	&785	&62	&463\\
32 \& 21 &2484	&1608	&47	&852\\
26 \& 5 &173	&46	    &161 &	43\\
26 \& 10 &693	&315	&81	&208\\
26 \& 15 &1560	&932	&54	&519\\
26 \& 20 &2773	&1970	&40	&989\\
\bottomrule
\end{tabular}
\end{table}

\begin{table}
\centering
\caption{\label{preSN2}Roche Lobe radii. See text.}
\begin{tabular}{c c c }
\toprule
Conf. (M$_\odot$) &Sep. (R$_\odot$) & RL (R$_\odot$)  \\
\midrule
41 \& 64 &1938	&810   \\
32 \& 64 &1442	&634	 \\
32 \& 42 &2484	&1000	\\
26 \& 31 &2773	&1093 \\
\bottomrule
\end{tabular}
\end{table}
We calculated the pre-SN binary parameters for different companion and pre-SN progenitor masses, assuming the peculiar velocity of the runaway star equals the orbital velocity before the SN.
We found large binary separations (Table \ref{preSN}) 
compared to the confirmed runaway star HD~37424 inside S~147 \citep{2015MNRAS.448.3196D}.
HD~37424 shows a large peculiar velocity, and it is highly likely 
that the pre-SN system experienced a common envelope evolution 
and in the most compact case, the components were separated only by $\sim9$ R$_\odot$.
We calculated the pre-SN binary parameters for the pre-SN mass configurations 
according to Table \ref{mass} as well as considering smaller progenitor masses such as 15, 10 and 5 M$_\odot$ owing to
the possible further mass loss due to binary interactions 
\citep{1976IAUS...73...35V, 2000ARA&A..38..113T,2017ApJ...842..125Z}
(Table \ref{preSN}).
We then calculated the Roche Lobe radii for the initial progenitor masses but using
the largest binary separations in the previous calculations as if the separation did not change through the history of the system, and the mass loss is due to the 
metallicity and not the binary interaction (Table \ref{preSN2}).
The Roche Lobe radii we found in these cases are still large; however, a red supergiant can still fill in and accrete on the secondary.
On the other hand, the values are much larger than blue supergiant radii.
Common envelope evolution and a corresponding violent mass loss might not be likely. 
Yet, SNe of such massive stars are expected to form black holes by fallback.
The channel generating NSs from stars more massive than 25 M$_\odot$ 
is the extreme mass loss of the star during its main-sequence lifetime.
\cite{2017ApJ...842..125Z}
suggests that high metallicity alone is insufficient for such a mass loss; the primary star should experience mass loss through close binary interactions.  
We cannot predict the history of the pre-SN system; it is also possible that the separation got larger due to the mass transfer.
Detailed simulations are needed for a clearer view of the matter.
\section{Summary}

We detected a massive, B0.5-type runaway star inside SNR IC~443.
By mid-high resolution spectroscopy, we found that the star HD~254577 is a hot
($T_\mathrm{eff}=24000\pm1000$ K), and evolved
($\log(g~\rm[cm/s^2])=2.75\pm0.25$, $\xi=10$ km~s$^{-1}$) star with a projected rotational velocity of $v \sin i =150\pm20$ km~s$^{-1}$ and a 3-D peculiar velocity of $v_\mathrm{pec}=31.3^{+1.2}_{-0.9}$ km~s$^{-1}$.
The star's proper motion direction and the cometary tail of the NS imply that they have the same origin, a binary SN.
The NS transverse velocities (254-538 km~s$^{-1}$) calculated using the positions generated by the runaway star tracing back are consistent with typical pulsar velocities for reasonable SNR ages ($10-20$ kyr). 
Also, the high-velocity ISM absorption lines in the star's spectrum are shifting only toward the blue, suggesting that the star is inside the SNR, as expected from a runaway star linked to an SNR. 
We did not detect a significant variation in the RV of the star based on measurements from eight epochs; hence, it is most probably a single star as expected for a typical runaway star. 
Given the fact that as a hot, rarely found, isolated, fast-moving star,  HD~254577 is located inside the SNR IC~443 and moving away from its NS, it is the pre-SN binary companion to the progenitor of IC~443.

By studying the neighboring stars and the possible host open cluster, we found that HD~254577 had a ZAMS mass of $28-54$ M$_\odot$ and the ZAMS mass of the progenitor was even higher $31-64$ M$_\odot$.
We confirmed the previous high mass estimations in 
\cite{2018A&A...615A.157G} and
\cite{2008A&A...485..777T}
The distance of the SNR is the distance of the runaway star, $r_\mathrm{geo}=1701^{+55}_{-54}$ pc, which is consistent with the distance of $1.80\pm0.05$ kpc proposed by \cite{2020ApJ...891..137Z} based on the extinction toward the background stars.
The explosion site we determined is primarily separated from the geometrical center, supporting the highly asymmetric expansion due to the surrounding molecular clouds demonstrated by \cite{2021A&A...649A..14U}.
By studying the optical filaments of the SNR, we showed that the true extension of the SNR is somewhat larger, and the shock wave has possibly lost its energy toward E and S.
We also constructed the pre-SN binary system and displayed the orbital parameters.
Binary interaction is not as strong as the previously confirmed case HD~37424 inside SNR S~147. 
The common envelope phase that helps the progenitor lose a large amount of mass and end up with an NS instead of a black hole might not have happened in this system.
\section{Acknowledgements}
This work was conceived by the late Oktay H. GUSE\.INOV (1938-2009).
We acknowledge his leading role in all stages of this work and other related topics. His contribution to Galactic Astrophysics and his scholarship will be greatly missed.
This work is based on observations obtained with telescopes of the University Observatory Jena, operated by the Astrophysical Institute of the Friedrich-Schiller-University Jena.
We appreciate the observational support from Daniel Wagner, Hartmut Gilbert, Kim Werner, and Anna Pannicke.  
This research is based on data collected at the Subaru Telescope, with run ID: S22A0154S, which is operated by the National Astronomical Observatory of Japan. 
We are honored and grateful for the opportunity of observing the Universe from Mauna Kea, which has cultural, historical, and natural significance in Hawaii.
This research has made use of the “Aladin sky atlas” developed at CDS, Strasbourg Observatory, France. We gratefully acknowledge the support of NASA and the contributors of the SkyView surveys. The Digitized Sky Survey was produced at the Space Telescope Science Institute under U.S. Government grant NAG W-2166. The images of these surveys are based on photographic data obtained using the Oschin-Schmidt Telescope on Palomar Mountain and the UK Schmidt Telescope. The plates were
processed into the present compressed digital form with the permission of these
institutions. The research presented in this paper has used data from the Canadian
Galactic Plane Survey, a Canadian project with international partners supported
by the Natural Sciences and Engineering Research Council. We also acknowledge
the ROSAT data archive provided by the Max-Planck-Institute for Extraterrestrial Physics, Garching, Germany.


\twocolumn
\bibliographystyle{aa}
\bibliography{c2hd25}
\begin{appendix}
\section{Long Table Contents}
There will be three tables in the appendix:
\begin{itemize}
    \item Table \ref{clustermembers} - Possible cluster members within 10 pc.
    \item Table \ref{Lines} - Line strengths of SNR spectra.
    \item Table \ref{parameters} - Physical parameters of the SNR.
\end{itemize}
\longtab[1]{\begin{longtable}{@{}l l l l l l l@{}}
\caption{\label{clustermembers}Cluster members.}\\
\toprule
Gaia DR3 Source	    
& r$_{geo}$(pc)        
& $\varpi$ (mas)  	      
& G (mag) 	     
& G$_{BP}-$G$_{RP}$ (mag) 
&$\mu_\alpha^*$ (mas~yr$^{-1}$) 	
& $\mu_\delta$ (mas~yr$^{-1}$)   \\
\midrule
\endfirsthead
\caption{Continued.}\\
\toprule
Gaia DR3 Source	    
& r$_{geo}$(pc)        
& $\varpi$ (mas)  	      
& G (mag) 	     
& G$_{BP}-$G$_{RP}$ (mag) 
&$\mu_\alpha^*$ (mas~yr$^{-1}$) 	
& $\mu_\delta$ (mas~yr$^{-1}$)   \\
\midrule
\endhead
\bottomrule
\endfoot
\bottomrule
\endlastfoot
3377027132948376704	& $1755^{+104}_{-88}    $    &$0.5262\pm0.0292$     &$14.945\pm0.003$  &$1.625\pm0.009$	&$0.392\pm0.033$	&$-1.678\pm0.023$  \\[3pt]
3377016721947681664	& $1618^{+51	}_{-45} $       &$0.5744\pm0.0201$     &$13.948\pm0.003$  &$0.981\pm0.007$	&$0.504\pm0.024$	&$-1.683\pm0.016$  \\[3pt]
3377026961149682432	& $1740^{+92	}_{-87} $       &$0.5332\pm0.0260$     &$14.620\pm0.003$  &$1.707\pm0.007$	&$0.259\pm0.029$	&$-1.760\pm0.020$  \\[3pt]
3377004215002923904	& $1743^{+75	}_{-52} $       &$0.5293\pm0.0180$     &$11.374\pm0.003$  &$0.661\pm0.007$	&$0.372\pm0.022$	&$-1.976\pm0.016$  \\[3pt]
3377030122245583616	& $1730^{+75	}_{-70} $       &$0.5349\pm0.0231$     &$14.533\pm0.003$  &$2.032\pm0.011$	&$0.269\pm0.026$	&$-1.833\pm0.018$  \\[3pt]
3377002978052355840	& $1710^{+122}_{-83}    $    &$0.5455\pm0.0319$     &$15.296\pm0.003$  &$1.097\pm0.008$	&$0.425\pm0.037$	&$-1.744\pm0.027$  \\[3pt]
3377022700542134144	& $1704^{+267}_{-165}   $    &$0.5822\pm0.0568$     &$16.274\pm0.003$  &$1.707\pm0.015$	&$0.391\pm0.064$	&$-2.045\pm0.043$  \\[3pt]
3377023765694007040	& $1700^{+98	}_{-100}$       &$0.5465\pm0.0412$     &$15.471\pm0.003$  &$1.550\pm0.008$	&$0.391\pm0.043$	&$-2.084\pm0.030$  \\[3pt]
3377024074931648512	& $1758^{+155}_{-138}   $    &$0.5346\pm0.0440$     &$15.817\pm0.003$  &$1.496\pm0.009$	&$0.303\pm0.047$	&$-1.818\pm0.033$  \\[3pt]
3377002840613390592	& $1624^{+40	}_{-44} $       &$0.5694\pm0.0156$     &$13.496\pm0.003$  &$0.706\pm0.007$	&$0.964\pm0.019$	&$-2.524\pm0.014$  \\[3pt]
3377003076832759936	& $1784^{+69	}_{-61} $       &$0.5174\pm0.0248$     &$14.753\pm0.003$  &$1.339\pm0.007$	&$0.347\pm0.028$	&$-0.829\pm0.020$  \\[3pt]
3376999709578280064	& $1725^{+87	}_{-79} $       &$0.5511\pm0.0308$     &$ 9.601\pm0.003$  &$0.668\pm0.008$	&$0.400\pm0.036$	&$-1.812\pm0.025$  \\[3pt]
3376999713877203456	& $1831^{+148}_{-114}   $    &$0.5024\pm0.0367$     &$15.619\pm0.003$  &$1.258\pm0.010$	&$0.329\pm0.041$	&$-1.845\pm0.030$  \\[3pt]
3377029950446864896	& $1699^{+43	}_{-44} $       &$0.5506\pm0.0167$     &$13.052\pm0.003$  &$1.386\pm0.007$	&$0.195\pm0.019$	&$-2.280\pm0.015$  \\[3pt]
3377003252930240000	& $1817^{+144}_{-132}   $    &$0.5103\pm0.0383$     &$15.716\pm0.003$  &$1.447\pm0.009$	&$0.178\pm0.044$	&$-1.591\pm0.032$  \\[3pt]
3377054620738994816	& $1770^{+85	}_{-66} $       &$0.5259\pm0.0269$     &$13.962\pm0.003$  &$1.467\pm0.007$	&$0.470\pm0.025$	&$-1.959\pm0.020$  \\[3pt]
3377025792918552960	& $1627^{+132}_{-134}   $    &$0.5712\pm0.0502$     &$16.042\pm0.003$  &$1.549\pm0.013$	&$0.134\pm0.056$	&$-2.250\pm0.041$  \\[3pt]
3377010812072647424	& $1832^{+100}_{-117}   $    &$0.5052\pm0.0305$     &$15.303\pm0.003$  &$1.458\pm0.008$	&$0.347\pm0.035$	&$-1.565\pm0.026$  \\[3pt]
3377010537194741888	& $1840^{+86	}_{-110}$       &$0.5103\pm0.0267$     &$14.977\pm0.003$  &$1.421\pm0.010$	&$0.273\pm0.031$	&$-2.496\pm0.023$  \\[3pt]
3377010361097396864	& $1806^{+100}_{-112}   $    &$0.5139\pm0.0290$     &$14.995\pm0.003$  &$1.227\pm0.007$	&$0.249\pm0.034$	&$-2.265\pm0.026$  \\[3pt]
3377024899565349760	& $1762^{+71	}_{-62} $       &$0.5298\pm0.0240$     &$14.594\pm0.003$  &$1.040\pm0.007$	&$0.003\pm0.027$	&$-2.141\pm0.020$  \\[3pt]
3377026411393808768	& $1720^{+61	}_{-63} $       &$0.5438\pm0.0197$     &$13.042\pm0.003$  &$0.865\pm0.007$	&$0.516\pm0.019$	&$-1.723\pm0.013$  \\[3pt]
3377012938077937280	& $1731^{+115}_{-97}    $    &$0.5383\pm0.0342$     &$15.360\pm0.003$  &$1.695\pm0.010$	&$0.557\pm0.038$	&$-1.832\pm0.028$  \\[3pt]
3376810529157870336	& $1787^{+132}_{-96}    $    &$0.5147\pm0.0408$     &$15.804\pm0.003$  &$1.256\pm0.010$	&$0.758\pm0.047$	&$-1.468\pm0.037$  \\[3pt]
3377012731919116416	& $1579^{+64	}_{-54} $       &$0.5998\pm0.0234$     &$10.168\pm0.003$  &$0.908\pm0.028$	&$0.483\pm0.029$	&$-1.813\pm0.021$  \\[3pt]
3377009815640222464	& $1683^{+119}_{-193}   $    &$0.5865\pm0.0545$     &$16.366\pm0.003$  &$1.519\pm0.019$	&$0.495\pm0.060$	&$-1.739\pm0.045$  \\[3pt]
3377013148534813056	& $1699^{+85	}_{-68} $       &$0.5503\pm0.0292$     &$15.161\pm0.003$  &$1.042\pm0.007$	&$0.496\pm0.033$	&$-2.280\pm0.025$  \\[3pt]
3377055788970045952	& $1653^{+157}_{-114}   $    &$0.5646\pm0.0417$     &$15.823\pm0.003$  &$1.620\pm0.009$	&$0.495\pm0.043$	&$-1.433\pm0.034$  \\[3pt]
3377012697559377664	& $1704^{+42	}_{-37} $       &$0.5553\pm0.0139$     &$12.514\pm0.003$  &$0.750\pm0.007$	&$0.376\pm0.016$	&$-2.221\pm0.012$  \\[3pt]
3377011911584237824	& $1716^{+48	}_{-43} $       &$0.5403\pm0.0152$     &$13.240\pm0.003$  &$0.820\pm0.007$	&$0.436\pm0.017$	&$-1.564\pm0.014$  \\[3pt]
3377049294979539712	& $1640^{+88	}_{-92} $       &$0.5652\pm0.0373$     &$14.929\pm0.003$  &$1.051\pm0.008$	&$0.609\pm0.036$	&$-1.377\pm0.029$  \\[3pt]
3377012701858219392	& $1637^{+92	}_{-70} $       &$0.5713\pm0.0265$     &$14.910\pm0.003$  &$0.952\pm0.007$	&$0.416\pm0.031$	&$-1.759\pm0.023$  \\[3pt]
3377005142715976832	& $1617^{+37	}_{-39} $       &$0.5744\pm0.0147$     &$13.109\pm0.003$  &$1.882\pm0.007$	&$0.013\pm0.018$	&$-2.548\pm0.013$  \\[3pt]
3377011739785552896	& $1805^{+167}_{-154}   $    &$0.5352\pm0.0511$     &$16.222\pm0.003$  &$2.039\pm0.014$	&$0.407\pm0.051$	&$-1.737\pm0.040$  \\[3pt]
3377013182894545536	& $1767^{+59	}_{-61} $       &$0.5269\pm0.0179$     &$13.892\pm0.003$  &$0.804\pm0.007$	&$0.351\pm0.020$	&$-1.714\pm0.015$  \\[3pt]
3377004936557546752	& $1649^{+115}_{-126}   $    &$0.5873\pm0.0482$     &$16.246\pm0.003$  &$1.352\pm0.020$	&$0.309\pm0.060$	&$-1.420\pm0.045$  \\[3pt]
3377006684605372672	& $1648^{+182}_{-102}   $    &$0.5922\pm0.0487$     &$16.086\pm0.003$  &$1.438\pm0.013$	&$0.331\pm0.051$	&$-1.795\pm0.040$  \\[3pt]
3377013178596140800	& $1827^{+160}_{-145}   $    &$0.5036\pm0.0443$     &$15.566\pm0.003$  &$1.379\pm0.009$	&$0.361\pm0.042$	&$-1.713\pm0.033$  \\[3pt]
3377011877224494720	& $1730^{+111}_{-87}    $    &$0.5326\pm0.0328$     &$15.383\pm0.003$  &$1.355\pm0.008$	&$0.540\pm0.035$	&$-1.534\pm0.027$  \\[3pt]
3377049737357365760	& $1724^{+61	}_{-53} $       &$0.5511\pm0.0220$     &$12.690\pm0.003$  &$0.835\pm0.008$	&$0.109\pm0.022$	&$-1.498\pm0.017$  \\[3pt]
3377013316036365440	& $1702^{+70	}_{-65} $       &$0.5468\pm0.0303$     &$14.634\pm0.003$  &$0.900\pm0.008$	&$0.352\pm0.029$	&$-1.710\pm0.024$  \\[3pt]
3377008200732505600	& $1805^{+160}_{-169}   $    &$0.5254\pm0.0472$     &$15.887\pm0.003$  &$1.522\pm0.012$	&$0.442\pm0.053$	&$-1.702\pm0.040$  \\[3pt]
3377049363699001856	& $1776^{+105}_{-66}    $    &$0.5263\pm0.0274$     &$14.795\pm0.003$  &$0.941\pm0.007$	&$0.440\pm0.027$	&$-1.705\pm0.022$  \\[3pt]
3377049874797080448	& $1812^{+119}_{-105}   $    &$0.5249\pm0.0348$     &$15.368\pm0.003$  &$1.382\pm0.008$	&$0.486\pm0.035$	&$-1.675\pm0.028$  \\[3pt]
3377037230412601472	& $1793^{+71	}_{-78} $       &$0.5287\pm0.0184$     &$12.874\pm0.003$  &$0.664\pm0.009$	&$0.381\pm0.018$	&$-1.527\pm0.015$  \\[3pt]
3377049363698996992	& $1762^{+109}_{-110}   $    &$0.5208\pm0.0382$     &$15.672\pm0.003$  &$1.333\pm0.014$	&$0.703\pm0.040$	&$-1.218\pm0.034$  \\[3pt]
3377052146837781120	& $1702^{+101}_{-86}    $    &$0.5395\pm0.0313$     &$14.980\pm0.003$  &$1.011\pm0.008$	&$0.427\pm0.031$	&$-1.687\pm0.026$  \\[3pt]
3377052142539900416	& $1648^{+128}_{-132}   $    &$0.5726\pm0.0424$     &$15.691\pm0.003$  &$1.533\pm0.008$	&$0.468\pm0.043$	&$-1.653\pm0.033$  \\[3pt]
3377049810375579264	& $1723^{+160}_{-148}   $    &$0.5583\pm0.0509$     &$15.963\pm0.003$  &$1.458\pm0.010$	&$0.201\pm0.054$	&$-2.289\pm0.041$  \\[3pt]
3377051356563802752	& $1733^{+131}_{-144}   $    &$0.5585\pm0.0449$     &$15.646\pm0.003$  &$1.363\pm0.009$	&$0.239\pm0.045$	&$-1.438\pm0.035$  \\[3pt]
3376810116840976000	& $1831^{+110}_{-90}    $    &$0.5032\pm0.0294$     &$15.113\pm0.003$  &$0.946\pm0.009$	&$0.385\pm0.035$	&$-1.749\pm0.026$  \\[3pt]
3377037647028208256	& $1753^{+99	}_{-99} $       &$0.5364\pm0.0336$     &$15.337\pm0.003$  &$1.367\pm0.009$	&$0.571\pm0.034$	&$-1.048\pm0.027$  \\[3pt]
3377051287844324864	& $1747^{+120}_{-94}    $    &$0.5337\pm0.0351$     &$15.257\pm0.003$  &$1.238\pm0.009$	&$0.450\pm0.035$	&$-1.668\pm0.027$  \\[3pt]
3377036100840002176	& $1717^{+67	}_{-57} $       &$0.5402\pm0.0209$     &$13.836\pm0.003$  &$0.806\pm0.008$	&$0.289\pm0.018$	&$-2.276\pm0.015$  \\[3pt]
3377036100840001408	& $1725^{+61	}_{-62} $       &$0.5399\pm0.0203$     &$13.262\pm0.003$  &$0.784\pm0.007$	&$0.261\pm0.017$	&$-2.452\pm0.015$  \\[3pt]
3377051455346018176	& $1746^{+53	}_{-66} $       &$0.5341\pm0.0203$     &$14.086\pm0.003$  &$1.128\pm0.008$	&$0.414\pm0.021$	&$-1.709\pm0.016$  \\[3pt]
3377051047326151680	& $1788^{+80	}_{-80} $       &$0.5177\pm0.0282$     &$14.973\pm0.003$  &$1.339\pm0.010$	&$0.419\pm0.029$	&$-1.664\pm0.023$  \\[3pt]
3377007273019751680	& $1665^{+152}_{-113}   $    &$0.5626\pm0.0546$     &$16.476\pm0.003$  &$1.889\pm0.021$	&$0.467\pm0.063$	&$-1.750\pm0.048$  \\[3pt]
3377036062191097472	& $1606^{+157}_{-126}   $    &$0.5979\pm0.0572$     &$16.316\pm0.003$  &$1.662\pm0.014$	&$0.388\pm0.053$	&$-1.623\pm0.044$  \\[3pt]
3377036547516582784	& $1634^{+134}_{-98}    $    &$0.5817\pm0.0437$     &$15.929\pm0.003$  &$1.334\pm0.010$	&$0.426\pm0.044$	&$-2.057\pm0.035$  \\[3pt]
3377007650976873216	& $1768^{+214}_{-135}   $    &$0.5332\pm0.0533$     &$16.414\pm0.003$  &$1.663\pm0.014$	&$0.358\pm0.061$	&$-1.772\pm0.046$  \\[3pt]
3377051871959852032	& $1786^{+96	}_{-94} $       &$0.5146\pm0.0299$     &$14.831\pm0.003$  &$1.227\pm0.008$	&$0.475\pm0.028$	&$-1.652\pm0.023$  \\[3pt]
3377039326357346816	& $1703^{+77	}_{-76} $       &$0.5579\pm0.0270$     &$14.934\pm0.003$  &$0.861\pm0.008$	&$0.114\pm0.029$	&$-0.522\pm0.023$  \\[3pt]
3377039433734571520	& $1668^{+55	}_{-55} $       &$0.5581\pm0.0191$     &$13.736\pm0.003$  &$1.036\pm0.007$	&$0.480\pm0.018$	&$-1.658\pm0.015$  \\[3pt]
3376818260098920192	& $1881^{+168}_{-141}   $    &$0.5052\pm0.0493$     &$16.234\pm0.003$  &$1.405\pm0.013$	&$0.529\pm0.058$	&$-1.829\pm0.044$  \\[3pt]
3377052009398795264	& $1786^{+115}_{-112}   $    &$0.5160\pm0.0358$     &$15.191\pm0.003$  &$1.306\pm0.008$	&$0.331\pm0.034$	&$-1.733\pm0.028$  \\[3pt]
3377031698495039872	& $1723^{+110}_{-105}   $    &$0.5498\pm0.0444$     &$15.756\pm0.003$  &$1.391\pm0.009$	&$0.480\pm0.042$	&$-1.746\pm0.035$  \\[3pt]
3377058018057226496	& $1795^{+149}_{-137}   $    &$0.5190\pm0.0374$     &$15.346\pm0.003$  &$1.059\pm0.009$	&$0.486\pm0.036$	&$-1.669\pm0.029$  \\[3pt]
3377038334222942976	& $1807^{+184}_{-110}   $    &$0.5257\pm0.0433$     &$15.781\pm0.003$  &$1.238\pm0.009$	&$0.121\pm0.040$	&$-2.039\pm0.034$  \\[3pt]
3377057644395874944	& $1757^{+138}_{-133}   $    &$0.5447\pm0.0462$     &$15.816\pm0.003$  &$1.492\pm0.012$	&$0.377\pm0.044$	&$-1.574\pm0.036$  \\[3pt]
3377057747475083648	& $1768^{+56	}_{-50} $       &$0.5226\pm0.0208$     &$13.976\pm0.003$  &$0.896\pm0.007$	&$0.473\pm0.020$	&$-1.665\pm0.016$  \\[3pt]
3377038849619004032	& $1830^{+132}_{-144}   $    &$0.5067\pm0.0437$     &$15.630\pm0.003$  &$1.282\pm0.011$	&$0.484\pm0.038$	&$-1.757\pm0.031$  \\[3pt]
3377044690774512896	& $1782^{+55	}_{-47} $       &$0.5358\pm0.0157$     &$12.431\pm0.003$  &$0.873\pm0.007$	&$0.411\pm0.014$	&$-1.700\pm0.011$  \\[3pt]
3377046168243238528	& $1669^{+56	}_{-56} $       &$0.5549\pm0.0217$     &$14.120\pm0.003$  &$1.100\pm0.008$	&$0.331\pm0.022$	&$-1.849\pm0.017$  \\[3pt]
3377033936176712192	& $1768^{+54	}_{-57} $       &$0.5213\pm0.0155$     &$13.111\pm0.003$  &$0.882\pm0.007$	&$0.518\pm0.015$	&$-1.677\pm0.012$  \\[3pt]
3376842346275448448	& $1749^{+124}_{-110}   $    &$0.5465\pm0.0423$     &$15.694\pm0.003$  &$1.310\pm0.010$	&$0.429\pm0.042$	&$-1.883\pm0.034$  \\[3pt]
3377041392239630080	& $1751^{+127}_{-147}   $    &$0.5278\pm0.0483$     &$15.999\pm0.003$  &$1.722\pm0.009$	&$0.232\pm0.047$	&$-1.024\pm0.037$  \\[3pt]
3377041117361721728	& $1591^{+135}_{-115}   $    &$0.5974\pm0.0406$     &$15.650\pm0.003$  &$1.072\pm0.010$	&$0.273\pm0.039$	&$-2.225\pm0.031$  \\[3pt]
3377041735836992896	& $1664^{+92	}_{-98} $       &$0.5645\pm0.0380$     &$15.279\pm0.003$  &$1.249\pm0.008$	&$0.445\pm0.036$	&$-1.924\pm0.028$  \\[3pt]
3377041770196723968	& $1650^{+155}_{-146}   $    &$0.5737\pm0.0486$     &$15.769\pm0.003$  &$1.432\pm0.010$	&$0.328\pm0.053$	&$-2.075\pm0.040$  \\[3pt]
3377040464526993920	& $1715^{+48	}_{-33} $       &$0.5430\pm0.0161$     &$11.893\pm0.003$  &$1.036\pm0.007$	&$0.526\pm0.015$	&$-1.459\pm0.012$  \\[3pt]
3377040636325684992	& $1584^{+123}_{-112}   $    &$0.5913\pm0.0424$     &$15.747\pm0.003$  &$1.440\pm0.009$	&$0.640\pm0.044$	&$-1.127\pm0.034$  \\[3pt]
3376852070081305344	& $1752^{+143}_{-118}   $    &$0.5417\pm0.0381$     &$15.147\pm0.003$  &$1.520\pm0.008$	&$0.732\pm0.033$	&$-2.243\pm0.026$  \\[3pt]
3376841521642369536	& $1825^{+120}_{-128}   $    &$0.5199\pm0.0446$     &$15.680\pm0.003$  &$1.256\pm0.009$	&$0.499\pm0.047$	&$-1.657\pm0.035$  \\[3pt]
3376838704143150976	& $1783^{+126}_{-100}   $    &$0.5204\pm0.0403$     &$15.709\pm0.003$  &$1.809\pm0.010$	&$0.524\pm0.042$	&$-2.430\pm0.031$  \\[3pt]
\end{longtable}}
\longtab[2]{\begin{longtable}{@{}llcccccccc@{}}
\caption{\label{Lines} Line fluxes $F$ relative to H$\alpha$. The line ratios are presented.}\\
\toprule
RA 
& Dec
& H$\beta$ 
& $[$O$~${\sc iii}$]$ 
& $[$O$~${\sc iii}$]$ 
& $[$N$~${\sc ii}$]$ 
& H$\alpha$ 
& $[$N$~${\sc ii}$]$ 
& $[$S$~${\sc ii}$]$ 
& $[$S$~${\sc ii}$]$ \\
$\text{[hh mm ss.ss]}$
& [dd mm ss.s]
& ($\lambda$4861) 
& ($\lambda$4959) 
& ($\lambda$5007) 
& ($\lambda$6548) 
& ($\lambda$6563) 
& ($\lambda$6584) 
& ($\lambda$6716) 
& ($\lambda$6731) \\
\midrule
\endfirsthead
\caption{Continued.}\\
\toprule
RA & Dec
& H$\beta$ 
& $[$O$~${\sc iii}$]$ 
& $[$O$~${\sc iii}$]$ 
& $[$N$~${\sc ii}$]$ 
& H$\alpha$
& $[$N$~${\sc ii}$]$ 
& $[$S$~${\sc ii}$]$ 
& $[$S$~${\sc ii}$]$ \\
$\text{[hh mm ss.ss]}$
& [dd mm ss.s]
& ($\lambda$4861) 
& ($\lambda$4959) 
& ($\lambda$5007) 
& ($\lambda$6548) 
& ($\lambda$6563) 
& ($\lambda$6584) 
& ($\lambda$6716) 
& ($\lambda$6731) \\
\midrule
\endhead
\bottomrule
\endfoot
\bottomrule
\endlastfoot
06:15:20.98 & +22:33:50.1 & 25$\pm10$ & $-$ & 78$\pm9$ & 24$\pm3$ & 100$\pm25$&  65$\pm5$ & 61$\pm9$ & 50$\pm7$ \\
06:15:23.08 &+22:33:33.9 & 43$\pm2$ & 27$\pm4$ & 90$\pm9$ & 23$\pm7$ & 100$\pm15$&  62$\pm4$ & 50$\pm9$ & 43$\pm2$ \\
06:15:24.40 &+22:33:08.8 & 21$\pm6$ & 31$\pm5$ & 88$\pm9$ & 22$\pm6$ & 100$\pm7$&  64$\pm9$ & 51$\pm3$ & 43$\pm3$ \\
06:15:24.40 &+22:33:08.8 & 18$\pm12$ & 25$\pm4$ & 77$\pm5$ & 23$\pm12$ & 100$\pm15$&  62$\pm6$ & 52$\pm10$ & 44$\pm9$ \\
06:15:24.40 &+22:33:08.8 & 10$\pm3$ & 13$\pm4$ & 39$\pm7$ & 24$\pm7$ & 100$\pm16$&  65$\pm7$ & 58$\pm6$ & 49$\pm4$ \\
06:15:24.40 &+22:33:08.8 & 19$\pm9$ & 27$\pm9$ & 79$\pm17$ & 28$\pm19$ & 100$\pm17$&  64$\pm2$ & 53$\pm16$ & 43$\pm13$ \\
06:15:24.40 &+22:33:08.8 & 13$\pm2$ & $-$ & 41$\pm1$ & 19$\pm1$ & 100$\pm2$&  68$\pm1$ & 59$\pm2$ & 48$\pm2$ \\
06:15:24.40 &+22:33:08.8 & 14$\pm1$ & 10$\pm1$ & 30$\pm1$ & 20$\pm3$ & 100$\pm2$&  63$\pm2$ & 55$\pm2$ & 46$\pm2$ \\
06:15:24.40 &+22:33:08.8 & 16$\pm1$ & 13$\pm2$ & 36$\pm1$ & 18$\pm2$ & 100$\pm2$&  70$\pm2$ & 59$\pm4$ & 48$\pm3$ \\
06:15:24.40 &+22:33:08.8 & 19$\pm1$ & 14$\pm1$ & 45$\pm2$ & 21$\pm2$ & 100$\pm3$&  62$\pm2$ & 54$\pm2$ & 46$\pm1$ \\
06:15:24.40 &+22:33:08.8 & 19$\pm1$ & 14$\pm2$ & 41$\pm1$ & 18$\pm2$ & 100$\pm2$&  61$\pm2$ & 54$\pm3$ & 45$\pm2$ \\
06:15:24.40 &+22:33:08.8 & 16$\pm1$ & 16$\pm1$ & 37$\pm1$ & 19$\pm2$ & 100$\pm2$&  63$\pm2$ & 56$\pm1$ & 48$\pm2$ \\
06:15:24.40 &+22:33:08.8 &  $-$ &  $-$ &  $-$ & 18$\pm2$ & 100$\pm2$&  64$\pm2$ & 57$\pm2$ & 49$\pm3$ \\
06:15:36.60 &+22:15:25.0 &  $-$ &  141$\pm3$ &  $-$ & 18$\pm2$ & 100$\pm2$&  69$\pm1$ & 67$\pm1$ & 46$\pm2$ \\
06:15:56.00 &+22:25:43.2 &  20$\pm6$ & 15$\pm4$ &  42$\pm5$ & 26$\pm16$ & 100$\pm18$&  61$\pm6$ & 76$\pm9$ & 55$\pm12$ \\
06:16:01.67 &+22:26:14.5 & 27$\pm7$ & 13$\pm5$ & 35$\pm5$ & 17$\pm9$ & 100$\pm23$&  63$\pm16$ & 52$\pm6$ & 43$\pm9$ \\
06:16:09.52 &+22:40:37.9 & $-$ &  $-$ &  $-$ & 44$\pm9$ & 100$\pm18$&  93$\pm11$ & 91$\pm11$ & 58$\pm4$ \\
06:16:16.88 &+22:26:06.6 & 23$\pm8$ &  19$\pm3$ &  55$\pm2$ & 21$\pm10$ & 100$\pm6$&  60$\pm5$ & 59$\pm5$ & 45$\pm5$ \\
06:16:20.63 &+22:48:04.7 & $-$ &  $-$ &  $-$ & 16$\pm7$ & 100$\pm18$&  67$\pm21$ & 65$\pm9$ & 55$\pm11$ \\
06:16:23.82 &+22:48:13.5 & 7$\pm3$ &  7$\pm4$ &  18$\pm16$ & 26$\pm7$ & 100$\pm9$&  87$\pm9$ & 76$\pm9$ & 61$\pm5$ \\
06:16:26.26 &+22:48:10.7 & $-$ & 28$\pm11$ &  73$\pm9$ & 18$\pm6$ & 100$\pm11$&  18$\pm8$ & 66$\pm5$ & 54$\pm10$ \\
06:16:30.64 &+22:25:36.7 & 15$\pm6$ & $-$ &  10$\pm4$ & 17$\pm10$ & 100$\pm22$&  61$\pm29$ & 77$\pm8$ & 51$\pm7$ \\
06:16:33.69 &+22:37:27.9 & $-$ & $-$ &  $-$ & 18$\pm13$ & $-$ &  45$\pm8$ & 65$\pm9$ & 48$\pm6$ \\
06:16:33.84 &+22:43:05.4 & $-$ & $-$ &  $-$ & 102$\pm12$ & 100$\pm8$&  81$\pm3$ & 76$\pm4$ & 54$\pm9$ \\
06:16:36.56 &+22:46:58.2 & 24$\pm3$ & $-$ &  $-$ & 49$\pm4$ & 100$\pm3$&  103$\pm2$ & 114$\pm4$ & 81$\pm3$ \\
06:16:36.83 &+22:01:41.9 & 24$\pm3$ & 16$\pm2$ &  50$\pm2$ & 16$\pm2$ & 100$\pm8$&  51$\pm2$ & 60$\pm3$ & 45$\pm2$ \\
06:16:48.47 &+22:23:22.8 & $-$ & $-$ &  $-$ & $-$ & 100$\pm3$&  46$\pm2$ & 101$\pm3$ & 64$\pm2$ \\
06:16:49.49 &+22:08:37.1 & 32$\pm3$ & $-$ &  9$\pm2$ & 20$\pm3$ & 100$\pm6$&  55$\pm5$ & 76$\pm9$ & 55$\pm8$ \\
06:16:50.26 &+22:20:38.3 & 27$\pm3$ & $-$  &  8$\pm3$ & 16$\pm3$ & 100$\pm6$&  45$\pm5$ & 81$\pm8$ & 55$\pm2$ \\
06:16:51.81 &+22:18:52.6 & $-$ & $-$ &  $-$  & 27$\pm19$ & 100$\pm21$ &  60$\pm6$ & 79$\pm5$ & 58$\pm4$ \\
06:16:54.40 &+22:20:31.4 & 16$\pm1$ & 45$\pm2$ &  8$\pm2$ & 19$\pm4$ & 100$\pm5$&  55$\pm5$ & 87$\pm4$ & 65$\pm3$ \\
06:16:56.41 &+22:56:15.4 & $-$ & $-$ &  66$\pm12$ & 65$\pm14$ & 100$\pm21$&  $-$ & 48$\pm12$ & 39$\pm11$ \\
06:17:03.38 &+22:21:25.3 & $-$ & $-$ &  $-$ & 18$\pm3$ & 100$\pm9$ &  55$\pm11$ & 86$\pm5$ & 65$\pm4$ \\
06:17:08.16 &+22:28:00.8 & $-$ & $-$ &  $-$ & 23$\pm12$ & 100$\pm21$&  59$\pm15$ & 96$\pm11$ & 64$\pm11$ \\
06:17:09.22 &+22:29:58.8 & $-$ & $-$ & 35$\pm8$ &  15$\pm1$ & 100$\pm16$ & 58$\pm5$&  78$\pm6$ & 55$\pm5$ \\
06:17:13.41 &+22:11:43.3 & 26$\pm1$ & $-$ &  7$\pm1$ & 15$\pm3$ & 100$\pm5$&  46$\pm2$ & 81$\pm2$ & 55$\pm1$ \\
06:17:14.89 &+22:51:51.9 & $-$ & $-$ & $-$ & 22$\pm2$ & 100$\pm6$&  64$\pm1$ & 50$\pm5$ & 45$\pm6$ \\
06:17:15.12 &+22:43:00.9 & 42$\pm3$ & 26$\pm2$ & 76$\pm8$ & 18$\pm5$ & 100$\pm12$&  57$\pm4$ & 56$\pm2$ & 46$\pm7$ \\
06:17:20.39 &+22:22:15.7 & 26$\pm6$ & 5$\pm3$ & $-$ & 16$\pm7$ & 100$\pm9$&  48$\pm11$ & 84$\pm5$ & 63$\pm8$ \\
06:17:22.34 &+22:53:52.7 & 29$\pm3$ & 40$\pm9$ & 120$\pm5$ & 21$\pm6$ & 100$\pm8$&  70$\pm7$ & 57$\pm8$ & 53$\pm4$ \\
06:17:22.91 &+22:25:10.5 & $-$& $-$& $-$ & 9$\pm5$ & 100$\pm2$&  40$\pm2$ & 94$\pm10$ & 70$\pm11$ \\
06:17:22.94 &+22:47:27.8 & 22$\pm5$ & 9$\pm2$ & 21$\pm3$ & 20$\pm7$ & 100$\pm9$&  63$\pm6$ & 77$\pm9$ & 63$\pm13$ \\
06:17:25.27 &+22:32:42.3 & $-$ & $-$ &  $-$ & 23$\pm2$ & 100$\pm3$&  57$\pm2$ & 77$\pm3$ & 50$\pm3$ \\
06:17:26.48 &+22:24:02.2 & $-$ & $-$ &  $-$ & 25$\pm2$ & 100$\pm12$&  56$\pm10$ & 100$\pm13$ & 69$\pm14$ \\
06:17:26.48 &+22:24:02.2 & $-$ & $-$ &  $-$ & 22$\pm8$ & 100$\pm7$&  45$\pm15$ & 92$\pm12$ & 62$\pm25$ \\
06:17:27.44 &+22:09:15.8 & $-$ & $-$ &  $-$ & 18$\pm3$ & 100$\pm4$&  62$\pm2$ & 52$\pm2$ & 36$\pm2$ \\
06:17:29.80 &+22:50:17.5  & 35$\pm3$ & 10$\pm2$ & 32$\pm2$ & 19$\pm5$ & 100$\pm7$&  58$\pm5$ & 68$\pm4$ & 52$\pm5$ \\
06:17:32.47 &+22:26:49.3 & 16$\pm3$ & 3$\pm1$ & 9$\pm2$ & 14$\pm8$ & 100$\pm9$&  44$\pm9$ & 71$\pm5$ & 53$\pm6$ \\
06:17:35.02 &+22:52:09.6 & 30$\pm2$ & 11$\pm2$ & 32$\pm2$ & 16$\pm3$ & 100$\pm7$&  49$\pm6$ & 56$\pm7$ & 56$\pm5$ \\
06:17:35.85 &+22:32:36.4 & 44$\pm14$ & 7$\pm2$ & 7$\pm5$ & 17$\pm9$ & 100$\pm12$&  53$\pm8$ & 58$\pm13$ & 42$\pm11$ \\
06:17:36.53 &+22:39:45.7 & 27$\pm1$ & 10$\pm1$ & 27$\pm3$ & 17$\pm2$ & 100$\pm12$&  56$\pm6$ & 82$\pm10$ & 60$\pm15$ \\
06:17:36.53 &+22:39:45.7 & 16$\pm3$ & 3$\pm1$ & 9$\pm2$ & 14$\pm8$ & 100$\pm9$&  44$\pm9$ & 71$\pm5$ & 53$\pm6$ \\
06:17:36.59 &+22:46:23.1 & 14$\pm1$ & 5$\pm1$ & 15$\pm2$ & 16$\pm1$ & 100$\pm18$&  47$\pm2$ & 58$\pm3$ & 44$\pm2$ \\
06:17:37.68 &+22:48:28.1 & $-$ & $-$ &  127$\pm17$ & 37$\pm6$ & 100$\pm15$&  76$\pm15$ & 83$\pm16$ & 65$\pm13$ \\
06:17:43.26 &+22:48:28.1 & $-$ & $-$ &  $-$ & 22$\pm2$ & 100$\pm2$&  61$\pm9$ & 90$\pm6$ & 68$\pm7$ \\
06:17:44.78 &+22:57:27.0 & 26$\pm5$ & 16$\pm2$ & 41$\pm8$ & 12$\pm5$ & 100$\pm16$&  38$\pm7$ & 21$\pm6$ & 14$\pm5$ \\
06:17:46.00 &+22:33:07.1 & $-$ & $-$ &  $-$ & 20$\pm3$ & 100$\pm3$&  64$\pm4$ & 101$\pm2$ & 74$\pm4$ \\
06:17:46.80 &+22:30:31.1 & $-$ & $-$ &  $-$ & 17$\pm2$ & 100$\pm5$&  67$\pm5$ & 98$\pm3$ & 73$\pm4$ \\
06:17:48.00 &+22:05:24.1 & $-$ & $-$ & $-$ & 18$\pm3$ & 100$\pm11$&  45$\pm6$ & 77$\pm8$ & 55$\pm6$ \\
06:17:48.05 &+22:49:07.2 & 102$\pm2$ & 23$\pm7$ & 71$\pm5$ & 22$\pm1$ & 100$\pm5$&  16$\pm4$ & 79$\pm6$ & 61$\pm4$ \\
06:17:49.70 &+22:01:25.3 & $-$ & $-$ &  $-$ & 43$\pm2$ & 100$\pm5$&  104$\pm4$ & 77$\pm3$ & 55$\pm3$ \\
06:17:51.52 &+22:42:55.7 & 11$\pm2$ & $-$ & 3$\pm1$ & 15$\pm2$ & 100$\pm3$&  51$\pm2$ & 83$\pm2$ & 63$\pm2$ \\
06:17:57.14 &+22:42:57.2 & $-$ & $-$ &  $-$ & 15$\pm4$ & 100$\pm8$&  42$\pm6$ & 73$\pm5$ & 55$\pm4$ \\
06:17:58.36 &+22:48:47.8 & 37$\pm2$ & 9$\pm3$ & 26$\pm2$ & 15$\pm3$ & 100$\pm4$&  50$\pm6$ & 60$\pm2$ & 51$\pm3$ \\
06:18:03.13 &+22:34:45.4 & 25$\pm4$ & $-$ & 7$\pm1$ & 12$\pm2$ & 100$\pm10$&  38$\pm6$ & 70$\pm9$ & 51$\pm6$ \\
06:18:03.43 &+22:51:02.6 & 10$\pm4$ & $-$ & 31$\pm5$ & 18$\pm7$ & 100$\pm8$&  48$\pm7$ & 33$\pm4$ & 39$\pm9$ \\
06:18:03.52 &+22:27:22.6  & 16$\pm2$ & $-$ & $-$ & 14$\pm4$ & 100$\pm6$&  44$\pm3$ & 69$\pm4$ & 52$\pm4$ \\
06:18:07.00 &+22:44:56.2 & 47$\pm2$ & 10$\pm3$ & 28$\pm2$ & 16$\pm4$ & 100$\pm6$&  51$\pm3$ & 78$\pm4$ & 63$\pm4$ \\
06:18:09.99 &+22:28:53.8 & 14$\pm1$ & $-$ & 3$\pm1$ & 14$\pm1$ & 100$\pm4$&  45$\pm2$ & 113$\pm5$ & 83$\pm3$ \\
06:18:11.00 &+22:49:34.1 & 16$\pm2$ & $-$ & $-$ & 14$\pm3$ & 100$\pm3$&  44$\pm3$ & 59$\pm3$ & 41$\pm3$ \\
06:18:11.30 &+22:31:14.1 & 44$\pm3$ & $-$ & 8$\pm1$ & 17$\pm6$ & 100$\pm3$&  52$\pm6$ & 78$\pm4$ & 57$\pm3$ \\
06:18:13.27 &+22:28:23.1 &  17$\pm2$ & $-$ & 3$\pm1$ & 12$\pm4$ & 100$\pm11$&  38$\pm9$ & 86$\pm7$ & 63$\pm6$ \\
06:18:16.42 &+22:41:56.4 & 29$\pm1$ & 3$\pm1$ & 11$\pm1$ & 16$\pm3$ & 100$\pm9$&  48$\pm5$ & 61$\pm4$ & 49$\pm5$ \\
06:18:17.17 &+22:39:05.7 & 38$\pm3$ & 4$\pm1$ & 13$\pm2$ & 15$\pm4$ & 100$\pm8$&  47$\pm8$ & 85$\pm5$ & 67$\pm7$ \\
06:18:22.88 &+22:45:08.4 & $-$ & $-$ & $-$ & 21$\pm4$ & 100$\pm9$&  39$\pm9$ & 41$\pm6$ & 41$\pm3$ \\
06:18:23.45 &+22:35:32.6  & $-$ & $-$ & $-$ & 14$\pm3$ & 100$\pm4$&  39$\pm3$ & 73$\pm4$ & 55$\pm3$ \\
06:18:23.99 &+22:25:59.1 & $-$ & $-$ & $-$ & 58$\pm4$ & 100$\pm4$&  $-$ & 48$\pm3$ & 32$\pm2$ \\
06:18:31.43 &+22:29:46.2 & $-$ & $-$ & $-$ & 28$\pm2$ & 100$\pm4$&  $-$ & 78$\pm20$ & 74$\pm30$ \\
06:18:31.51 &+22:29:46.3 & $-$ & $-$ & $-$ & 18$\pm1$ & 100$\pm4$&  44$\pm4$ & 72$\pm5$ & 54$\pm4$ \\
06:18:32.11 &+22:47:51.3 & 16$\pm2$ & 6$\pm2$ & 14$\pm5$ & 15$\pm2$ & 100$\pm2$&  41$\pm3$ & 42$\pm4$ & 28$\pm3$ \\
06:18:36.24	 &+22:32:48.3 & $-$ & $-$ & $-$ & $-$ & 100$\pm33$& 65$\pm17$ & 70$\pm18$ & 57$\pm17$ \\
06:18:36.50 &+22:33:39.0 & $-$ & $-$ & $-$ & 18$\pm2$ & 100$\pm3$& 73$\pm4$ & 70$\pm4$ & 43$\pm3$ \\
06:18:42.68 & +22:43:35.8 & 20$\pm4$ & $-$ & 13$\pm3$ & 19$\pm5$ & 100$\pm23$&  56$\pm8$ & 51$\pm11$ & 40$\pm9$ \\
06:18:45.77 &+22:39:18.5 & $-$ & $-$ &  $-$ & 17$\pm3$ & 100$\pm3$&  54$\pm8$ & 53$\pm6$ & 48$\pm12$ \\
06:18:46.87 &+22:57:40.3 & 49$\pm6$ & $-$ & 21$\pm3$ & 11$\pm3$ & 100$\pm7$&  40$\pm9$ & 24$\pm5$ & 16$\pm4$ \\
06:18:46.90 &+22:30:27.5 & 25$\pm4$ & $-$ & $-$ & 16$\pm3$ & 100$\pm4$&  43$\pm3$ & 56$\pm2$ & 42$\pm2$ \\
06:18:48.44 &+22:56:13.8 & 20$\pm4$ & 7$\pm1$ & 14$\pm1$ & 14$\pm2$ & 100$\pm5$&  47$\pm12$ & 34$\pm5$ & 24$\pm5$ \\
06:19:03.50 &+22:30:10.1 &  $-$ & $-$ & $-$ & 24$\pm3$ & 100$\pm5$&  51$\pm7$ & 50$\pm3$ & 37$\pm4$ \\
06:19:24.85 &+22:48:24.5 & 22$\pm2$ & 8$\pm2$ & 26$\pm3$ & 12$\pm2$ & 100$\pm10$&  42$\pm3$ & 33$\pm2$ & 23$\pm1$ \\
06:19:25.43 &+22:51:03.6 & 22$\pm2$ & 7$\pm2$ & 12$\pm4$ & 15$\pm3$ & 100$\pm12$&  39$\pm3$ & 27$\pm2$ & 19$\pm2$ \\
06:19:27.07 &+22:33:30.1 & $-$ & $-$ & $-$ & $-$  & 100$\pm7$&  51$\pm6$ & 37$\pm5$ & 24$\pm5$ \\
06:19:41.38 &+22:47:38.7 & $-$ & $-$ & $-$ & 15$\pm2$ & 100$\pm2$&  45$\pm1$ & 44$\pm1$ & 31$\pm2$ \\
06:19:43.10 &+22:45:26.7 & $-$ & $-$ & $-$ & 15$\pm2$ & 100$\pm2$&  56$\pm3$ & 47$\pm1$ & 37$\pm4$ \\
06:19:58.37 &+22:43:22.2 & $-$ & $-$ & $-$ & 16$\pm3$ & 100$\pm7$&  47$\pm4$ & 33$\pm3$ & 28$\pm3$ \\
\end{longtable}}
\longtab[3]{\begin{longtable}{l l l l l l l l l}
\caption{\label{parameters} The Physical parameters of the SNR.}\\
\toprule
RA 
&Dec
&$[$S$~${\sc ii}$]$/H$~\alpha$
& $N_{\rm e}$
& $[$O$~${\sc iii}$]$/H$\beta$
& $V_{\rm s}$   
& $N_{\rm H}$  
& E(B-V)
& $A_{\rm v}$\\
$\text{[hh mm ss.ss]}$
& [dd mm ss.s]
&
& (cm$^{-3}$)
&
& (km s$^{-1}$)
& ($10^{21}$ cm$^{-2}$)
& (mag)
& (mag) \\
\midrule
\endfirsthead
\caption{Continued.}\\
\toprule
RA 
&Dec
&$[$S$~${\sc ii}$]$/H$~\alpha$
& $N_{\rm e}$
& $[$O$~${\sc iii}$]$/H$\beta$
& $V_{\rm s}$  
& $N_{\rm H}$ 
& E(B-V)
& $A_{\rm v}$\\
$\text{[hh mm ss.ss]}$
& [dd mm ss.s]
&
& (cm$^{-3}$)
&
& (km s$^{-1}$)
& ($10^{21}$ cm$^{-2}$)
& (mag)
& (mag) \\
\midrule
\endhead
\bottomrule
\endfoot
\bottomrule
\endlastfoot
06:15:20.98 &+22:33:50.1 &1.12$\pm0.12$ &274$\pm9$ & 3.06$\pm0.09$ & 70 & 1.21 & 0.25$\pm0.06$ & 0.78$\pm0.41$ \\
06:15:23.08 &+22:33:33.9 &0.94$\pm0.03$ &363$\pm16$ & 2.74$\pm0.02$ & 70 & 1.22 & 0.22$\pm0.13$ & 0.69$\pm0.30$ \\
06:15:24.40 &+22:33:08.8 &0.96$\pm0.01$ &321$\pm12$ & 5.55$\pm0.03$ & 70 & 2.07 & 0.58$\pm0.21$ & 1.18$\pm0.42$ \\
06:15:24.40 &+22:33:08.8 &0.97$\pm0.05$ &302$\pm11$ & 5.80$\pm0.23$ & 70 & 2.07 & 0.55$\pm0.16$ & 1.71$\pm0.59$ \\
06:15:24.40 &+22:33:08.8 &1.08$\pm0.07$ &323$\pm16$ & 5.38$\pm0.41$ & 70 & 5.77 & 1.07$\pm0.29$ & 3.31$\pm0.91$ \\
06:15:24.40 &+22:33:08.8 &0.96$\pm0.13$ &268$\pm8$ & 5.59$\pm1.19$ & 70 & 2.61 & 0.48$\pm0.33$ & 1.50$\pm1.01$ \\
06:15:24.40 &+22:33:08.8 &1.08$\pm0.01$ &256$\pm7$ & 3.00$\pm0.09$ & 70 & 4.19 & 0.82$\pm0.13$ & 2.41$\pm1.05$ \\
06:15:24.40 &+22:33:08.8 &1.02$\pm0.01$ &306$\pm11$ & 2.78$\pm0.03$ & 70 & 3.89 & 0.72$\pm0.06$ & 2.23$\pm1.14$ \\
06:15:24.40 &+22:33:08.8 &1.09$\pm0.05$ &272$\pm9$ & 3.05$\pm0.15$ & 70 & 3.37 & 0.62$\pm0.05$ & 1.93$\pm0.94$ \\
06:15:24.40 &+22:33:08.8 &1.01$\pm0.01$ & 319$\pm12$ & 3.12$\pm0.05$ & 70 & 8.05 & 1.49$\pm0.79$ & 4.62$\pm2.08$ \\
06:15:24.40 &+22:33:08.8 & 1.00$\pm0.01$ &312$\pm12$ & 2.84$\pm0.01$ & 70 & 2.46 & 0.45$\pm0.05$ & 1.41$\pm0.81$ \\
06:15:24.40 &+22:33:08.8 &1.05$\pm0.01$ &357$\pm1$ & 3.19$\pm0.05$ & 70 & 3.19 & 0.59$\pm0.06$ & 1.83$\pm0.92$ \\
06:15:24.40 &+22:33:08.8 &1.08$\pm0.03$ &349$\pm1$ & - & - & - & - & - \\
06:15:36.60 &+22:15:25.0 &1.14$\pm0.01$ &38$\pm1$ & - & - & - & - & - \\
06:15:56.00 &+22:25:43.2 &1.32$\pm0.02$ &91$\pm1$ & 2.78$\pm0.05$ & 70 & 2.28 & 0.42$\pm0.13$ & 1.31$\pm0.85$ \\
06:16:01.67 &+22:26:14.5 &0.96$\pm0.06$ &296$\pm11$ & 1.76$\pm0.01$ & 70 & 9.22 & 0.97$\pm0.34$ & 0.53$\pm0.31$ \\
06:16:09.52 &+22:40:37.9 &1.50$\pm0.13$ &55$\pm1$ & - & - & - & - & - \\
06:16:16.88 &+22:26:06.6 &1.04$\pm0.04$ &144$\pm3$ & 3.20$\pm0.05$ & 70 & 1.66 & 0.56$\pm0.30$ & 0.96$\pm0.62$ \\
06:16:20.63 &+22:48:04.7 &1.20$\pm0.01$ &304$\pm11$ & - & - & - & - & - \\
06:16:23.82 &+22:48:13.5 &1.38$\pm0.02$ &223$\pm6$ & 3.43$\pm0.15$ & 70 & 7.17 & 1.32$\pm0.37$ & 4.11$\pm2.12$ \\
06:16:26.26 &+22:48:10.7 &1.20$\pm0.01$ &247$\pm7$ & - & - & - & - & - \\
06:16:30.64 &+22:25:36.7 &1.28$\pm0.13$ &4$\pm1$ & 0.28$\pm0.05$ & 80 & 2.03*$10^{20}$ & 0.32$\pm0.12$ & 0.12$\pm0.04$ \\
06:16:33.69 &+22:37:27.9 &1.13$\pm0.03$ &136$\pm2$ & - & - & - & - & - \\
06:16:33.84 &+22:43:05.4 &1.31$\pm0.02$ &78$\pm1$ & - & - & - & - & - \\
06:16:36.56 &+22:46:58.2 &1.95$\pm0.01$ &74$\pm1$ & - & - & - & - & - \\
06:16:36.89 &+22:01:41.9 &1.05$\pm0.03$ &141$\pm3$ & - & - & - & - & - \\
06:16:48.47	&+22:23:22.8 &1.65$\pm0.01$ &62$\pm1$ & - & - & - & - & - \\
06:16:49.49 &+22:08:37.1 &1.32$\pm0.09$ &93$\pm1$ & 0.29$\pm0.15$ & 80 & 1.85*$10^{20}$ & 0.30$\pm0.22$ & 0.11$\pm0.06$ \\
06:16:50.26 &+22:20:38.3 &1.35$\pm0.05$ &138$\pm3$ & 0.89$\pm0.10$ & 80 & 1.29 & 0.84$\pm0.32$ & 0.74$\pm0.35$\\
06:16:51.81 &+22:18:52.6 &1.37$\pm0.21$ &112$\pm2$ & - & - & - & - & - \\
06:16:54.40 &+22:20:31.4 &1.52$\pm0.01$ &108$\pm2$ & 0.17$\pm0.01$ & 80 & 1.40 & 0.56$\pm0.19$ & 0.80$\pm0.45$ \\
06:16:56.41 &+22:56:15.4 &0.87$\pm0.05$ &247$\pm3$ & - & - & - & - & - \\
06:17:03.38 &+22:21:25.3 &1.52$\pm0.05$ &131$\pm1$ & - & - & - & - & - \\
06:17:03.38 &+22:21:25.3 &1.47$\pm0.13$ &119$\pm1$ & - & - & - & - & - \\
06:17:08.16 &+22:28:00.8 &1.60$\pm0.11$ &7$\pm1$ & - & - & - & - & - \\
06:17:09.22 &+22:29:58.8 &1.34$\pm0.10$ &61$\pm1$ & - & - & - & - & - \\
06:17:13.41 &+22:11:43.3 &1.37$\pm0.30$ &8$\pm0.07$ & 0.28$\pm0.08$  & 80 & 1.05 & 0.91$\pm0.26$ & 0.60$\pm0.42$ \\
06:17:14.87 &+22:51:51.9 &0.95$\pm0.05$ &417$\pm21$ & - & - & - & - & - \\
06:17:15.12 &+22:43:00.9 &1.02$\pm0.03$ &267$\pm9$ & 2.41$\pm0.06$ & 70 & 1.13 & 0.91$\pm0.27$ & 0.65$\pm0.38$ \\
06:17:20.39 &+22:22:15.7 &1.47$\pm0.02$ &140$\pm12$ & - & - & - & - & - \\
06:17:22.34 &+22:53:52.7 &1.10$\pm0.03$ &462$\pm1$ & 5.67$\pm0.02$ & 70 & 7.72*$10^{20}$ & 0.79$\pm0.33$ & 0.44$\pm0.23$ \\
06:17:22.91 &+22:25:10.5 &1.64$\pm0.17$ &123$\pm2$ & - & - & - & - & - \\
06:17:22.94 &+22:47:27.8 &1.40$\pm0.03$ &276$\pm9$ & 1.36$\pm0.21$ & 80 & 1.98 & 0.37$\pm0.03$ & 1.38$\pm0.98$\\
06:17:25.27 &+22:32:42.3 &1.28$\pm0.04$ &21$\pm1$ & - & - & - & - & - \\
06:17:26.48 &+22:24:02.2 &1.70$\pm0.07$ &27$\pm1$ & - & - & - & - & - \\
06:17:26.48 &+22:24:02.2 &1.54$\pm0.26$ &4$\pm1$ & - & - & - & - & - \\
06:17:27.44 &+22:09:15.8 &0.88$\pm0.01$ &26$\pm1$ & - & - & - & - & - \\
06:17:29.80 &+22:50:17.5 &1.20$\pm0.02$ &152$\pm3$ & 1.26$\pm0.01$ & 80 & 1.53*$10^{20}$ & 0.43$\pm0.21$ & 0.09$\pm0.03$ \\
06:17:32.47 &+22:26:49.3 &1.24$\pm0.02$ &128$\pm2$ & 0.74$\pm0.07$ & 80 & 3.49 & 0.64$\pm0.17$ & 2.00$\pm1.02$ \\
06:17:35.02 &+22:52:09.6 &1.13$\pm0.04$ &684$\pm58$ & 1.45$\pm0.06$ & 70 & 5.06*$10^{20}$ & 0.09$\pm0.03$ & 0.29$\pm0.11$\\
06:17:35.85 &+22:32:36.4 &1.00$\pm0.11$ &99$\pm1$ & 0.33$\pm0.06$ & 80 & 1.24 & 0.93$\pm0.11$ & 0.71$\pm0.31$ \\
06:17:36.53 &+22:39:45.7 &1.41$\pm0.08$ &100$\pm1$ & 1.37$\pm0.11$ & 70 & 9.68*$10^{20}$ & 0.48$\pm0.13$ & 0.55$\pm0.32$ \\
06:17:36.59 &+22:46:23.1 &1.02$\pm0.21$ &154$\pm3$ & 1.41$\pm0.02$ & 70 & 3.84 & 0.71$\pm0.10$ & 2.20$\pm1.21$ \\
06:17:37.68 &+22:48:28.1 &1.48$\pm0.07$ &187$\pm4$ & - & - & - & - & - \\
06:17:43.26 &+22:48:28.2 &1.58$\pm0.11$ &145$\pm3$ & - & - & - & - & - \\
06:17:44.78 &+22:57:27.0 &0.35$\pm0.04$ &64$\pm1$ & 2.21$\pm0.07$ & 70 & 1.17 & 0.61$\pm0.32$ & 0.67$\pm0.36$ \\
06:17:46.00 &+22:33:07.1 &1.76$\pm0.08$ &88$\pm1$ & - & - & - & - & - \\
06:17:46.80 &+22:30:31.1 &1.72$\pm0.04$ &135$\pm3$ & - & - & - & - & - \\
06:17:48.00 &+22:05:24.1 & 1.32$\pm0.01$ & 50$\pm1$ & - & -& - & - & - \\
06:17:48.05 &+22:49:07.2 &1.41$\pm0.06$ &172$\pm4$ & 0.92$\pm0.01$ & 80 & 5.27 & 0.97$\pm0.03$ & 3.03$\pm1.81$ \\
06:17:49.70 &+22:01:25.3 &1.33$\pm0.01$ &80$\pm1$ & - & - & - & - & - \\
06:17:51.52 &+22:42:55.7 &1.46$\pm0.01$ &145$\pm1$ & 0.26$\pm0.09$ & 80 & 5.09 & 0.94$\pm0.16$ & 2.92$\pm1.73$ \\
06:17:57.14 &+22:46:57.2 &1.28$\pm0.02$ &149$\pm3$ & - & - & - & - & - \\
06:17:58.36 &+22:48:47.8  &1.12$\pm0.01$ &352$\pm15$ & 0.92$\pm0.08$ & 80 & 5.16 & 0.95$\pm0.05$ & 2.96$\pm1.21$ \\
06:18:03.13 &+22:34:45.4  &1.21$\pm0.02$ &129$\pm2$ & 0.28$\pm0.02$ & 80 & 1.27 & 0.23$\pm0.02$ & 0.73$\pm0.54$ \\
06:18:03.43 &+22:51:02.6  &0.73$\pm0.07$ &1257$\pm193$ & 3.05$\pm0.57$ & 70 & 5.57 & 1.03$\pm0.36$ & 3.20$\pm1.32$ \\
06:18:03.52 &+22:27:22.6 &1.21$\pm0.01$ &152$\pm3$ & - & - & - & - & - \\
06:18:07.00 &+22:44:56.2  &1.41$\pm0.02$ &241$\pm7$ & 0.82$\pm0.06$ & 80 & 1.59 & 0.92$\pm0.16$ & 0.91$\pm0.43$ \\
06:18:09.99 &+22:28:53.8  &1.97$\pm0.05$ &100$\pm2$ & 0.18$\pm0.04$ & 80 & 3.85 & 0.71$\pm0.07$ & 2.21$\pm1.51$ \\
06:18:11.00 &+22:49:34.1 &1.01$\pm0.02$ &54$\pm1$ & - & - & - & - & - \\
06:18:11.30 &+22:31:14.1 &1.35$\pm0.03$ &109$\pm2$ & 0.19$\pm0.02$ & 80 & 1.25 & 0.56$\pm0.26$ & 0.72$\pm0.37$ \\
06:18:13.27 &+22:28:23.1 &1.50$\pm0.03$ &90$\pm1$ & 0.14$\pm0.01$ & 80 & 3.03 & 0.56$\pm0.14$ & 1.74$\pm0.91$ \\
06:18:16.42 &+22:41:56.4 &1.11$\pm0.01$ &254$\pm8$ & 0.47$\pm0.03$ & 80 & 5.70*$10^{20}$ & 0.92$\pm0.42$ & 0.33$\pm0.11$ \\
06:18:17.17 &+22:39:05.7 &1.53$\pm0.01$ &204$\pm5$ & 0.19$\pm0.04$ & 80 & 6.89*$10^{20}$ & 0.83$\pm0.39$ & 0.39$\pm0.22$ \\
06:18:22.88 &+22:45:08.4 &0.83$\pm0.02$ &668$\pm54$ & - & - & - & - & - \\
06:18:23.45 &+22:35:32.6 &1.29$\pm0.01$ &149$\pm3$ & - & - & - & - & - \\
06:18:23.99 &+22:25:59.1 &0.79$\pm0.01$ &16$\pm1$ & - & - & - & - & - \\
06:18:31.43 &+22:29:46.2 &1.52$\pm0.55$ &552$\pm37$ & - & - & - & - & - \\
06:18:31.51 &+22:29:46.3 &1.26$\pm0.04$ &143$\pm3$ & - & - & - & - & -  \\
06:18:32.11 &+22:47:51.3 &0.71$\pm0.06$ &5$\pm0.02$ & 1.22$\pm0.29$ & 80 & 3.34 & 0.62$\pm0.11$ & 1.92$\pm0.89$ \\
06:18:36.24	 &+22:32:48.3 &1.27$\pm0.06$ &258$\pm8$ & - & - & - & - & -  \\
06:18:36.51 &+22:33:39.0 &1.13$\pm0.01$ &70$\pm6$ & - & - & - & - & - \\
06:18:42.68 & +22:43:35.8 &0.90$\pm0.01$ &170$\pm4$ & - & - & - & - & - \\
06:18:45.77 &+22:39:18.4 &1.02$\pm0.21$ &439$\pm24$ & - & - & - & - & - \\
06:18:46.87 &+22:57:40.3 &0.41$\pm0.08$ &8$\pm0.06$ & 0.43$\pm0.01$ & 80 & 1.87$\pm0.12$ &0.34$\pm0.09$ & - \\
06:18:46.90 &+22:30:27.5 &0.98$\pm0.33$ &120$\pm17$ & - & - & - &- & - \\
06:18:48.44 &+22:56:13.8 &0.58$\pm0.35$ &49$\pm29$ & 1.05$\pm0.02$ & 80 & 2.31 & 0.43$\pm0.02$ & 1.32$\pm0.78$ \\
06:19:03.50 &+22:30:10.1 &0.87$\pm0.03$ &101$\pm12$ & - & - & -  \\
06:19:24.85 &+22:48:24.5 &0.55$\pm0.03$ &48$\pm2$ & 1.49$\pm0.03$ & 70 & 1.79 &0.33$\pm0.01$ & 1.03$\pm0.56$ \\
06:19:25.43 &+22:51:03.6 &0.46$\pm0.01$ &34$\pm1$ & 0.89$\pm0.20$ & 80 & 1.94$\pm1$ & 0.36$\pm0.02$ & 1.11$\pm0.71$ \\
06:19:27.07 &+22:33:30.1 &0.62$\pm0.07$ &90$\pm1$ & - & - & -  \\
06:19:41.38 &+22:47:38.7 &0.75$\pm0.02$ &89$\pm1$ & - & - & -  \\
06:19:43.10 &+22:45:26.7 &0.84$\pm0.05$ &194$\pm5$ & - & - & - \\
06:19:58.37 &+22:43:22.2 &0.62$\pm0.02$ &313$\pm12$ & - & - & -  \\
\end{longtable}}
\end{appendix}

\label{LastPage}
\end{document}